\documentclass[12pt,letterpaper]{article}

\usepackage[utf8]{inputenc}
\usepackage[T1]{fontenc}
\usepackage{lmodern}
\usepackage[english]{babel}
\usepackage[margin=1.25in]{geometry}
\usepackage{microtype}
\usepackage{setspace}
\usepackage{amsmath,amsthm,amssymb,mathtools,latexsym}
\usepackage[font=small,labelfont=bf]{caption}
\usepackage{booktabs,graphicx,enumitem,subcaption,multirow,multicol,array,xcolor,mathrsfs,comment,float,tikz,adjustbox,tabularx}
\usetikzlibrary{arrows.meta,positioning,calc}
\usepackage[authoryear,round]{natbib}
\setcitestyle{citesep={;}}
\usepackage[hidelinks,bookmarksnumbered=true]{hyperref}
\hypersetup{
  pdftitle={Bang-Bang Correlation Sensitivity: A Local-to-Global Dichotomy for Correlation Preference},
  pdfauthor={Yi-Chun Chen; Soo Hong Chew; Xinhan Zhang}
}

\setlist[enumerate,1]{label=(\roman*)}

\theoremstyle{plain}
\newtheorem{thm}{Theorem}
\newtheorem{prop}{Proposition}
\newtheorem{lem}{Lemma}
\newtheorem{cor}{Corollary}

\theoremstyle{definition}
\newtheorem{defi}{Definition}
\newtheorem{axm}{Axiom}

\theoremstyle{remark}

\newenvironment{axmbis}[1]
  {\renewcommand{\theaxm}{\ref*{#1}*}%
   \renewcommand{\theHaxm}{fs.#1}%
   \addtocounter{axm}{-1}%
   \begin{axm}}
  {\end{axm}}

\newenvironment{thmbis}[1]
  {\renewcommand{\thethm}{\ref*{#1}*}
   \renewcommand{\theHthm}{fs.#1}
   \addtocounter{thm}{-1}
   \begin{thm}}
  {\end{thm}}

\newcommand{\supp}{\operatorname{supp}}
\newcommand{\one}{\mathbf 1}
\newcommand{\R}{\mathbb R}
\newcommand{\Sspace}{\mathcal S}
\newcommand{\Tspace}{\mathcal T}
\newcommand{\Pf}{\mathcal P_f}
\newcommand{\Jf}{\mathcal J_f}
\newcommand{\Mf}{\mathcal M_f}
\newcommand{\T}{\mathsf T}
\newcommand{\ip}[2]{\left\langle #1,#2\right\rangle}

\title{Bang-Bang Correlation Sensitivity: A Local-to-Global Dichotomy for Correlation Preference}
\author{
  Yi-Chun Chen\thanks{Department of Economics and Risk Management Institute,
    National University of Singapore.
    Email: \href{mailto:yichun@nus.edu.sg}{\texttt{yichun@nus.edu.sg}}.}
  \and
  Soo Hong Chew\thanks{Center for Intelligence Economic Science,
    Southwestern University of Finance and Economics; and Department of Economics,
    National University of Singapore.
    Email: \href{mailto:chew.soohong@gmail.com}{\texttt{chew.soohong@gmail.com}}.}
  \and
  Xinhan Zhang\thanks{Center for Intelligence Economic Science,
    Southwestern University of Finance and Economics.
    Email: \href{mailto:zhangxinhan@swufe.edu.cn}{\texttt{zhangxinhan@swufe.edu.cn}}.}
}
\date{}

\begin{document}
\hypersetup{pageanchor=false}
\begin{titlepage}
\begin{singlespace}
\maketitle
\end{singlespace}
\thispagestyle{empty}
\vspace{-1em}
\begin{abstract}
\begin{singlespace}
\noindent
Binary choice under risk can depend on how the alternative outcomes are jointly distributed. We show that in a finite outcome space, continuity, completeness, and betweenness impose a local-to-global restriction on this dependence: indifference between two alternatives with one common marginal distribution, however their outcomes are jointly distributed, implies correlation insensitivity for all lotteries supported on the same outcomes. In the correlation-sensitive case, these three axioms alone characterize correlation weighted utility, which combines statewise comparisons with a bilinear function of the marginal lotteries. In the correlation-insensitive case, an additional projective independence axiom yields Fishburn’s (1982) skew-symmetric bilinear representation. Strong independence reduces correlation weighted utility to Lanzani’s (2022) correlation expected utility representation. We extend the correlation-sensitive characterization to finitely supported lotteries on general metrizable outcome spaces. The resulting framework distinguishes correlation effects from departures from classical independence, and yields testable restrictions linking choices between identically distributed lotteries to correlated Allais behavior.
\end{singlespace}
\end{abstract}
\vspace{0.5em}
\begin{quote}
\small\singlespacing
\end{quote}
\vfill
\end{titlepage}
\setcounter{page}{1}
\hypersetup{pageanchor=true}

\section{Introduction}\label{section:intro}

Standard models of choice under risk identify alternatives with their marginal outcome distributions, making choice invariant to how their payoffs are jointly realized. In contrast, regret and salience theories allow correlation-sensitive choice through state-by-state comparisons of chosen and forgone payoffs \citep{bell1982regret,loomes1982regret,bordalo2012salience,lanzani2022correlation,koch2026counterfactual}. Experiments in psychology and finance document such sensitivity, even in comparisons between alternatives with identical marginal distributions \citep{diederich1999conflict,erev2023contradictory,bohren2024cognitive}. To illustrate, consider a problem involving two bets $C$ (``contrast'') and $L$ (``likelihood''), whose payoffs on three equally likely cards are


\begin{table}[h!]
\vspace{-1ex}
\begin{center}
\begin{tabular}{c@{\qquad}ccc}
\toprule
&Card 1&Card 2&Card 3\\
\midrule
$C$&0&50&100\\
$L$&50&100&0\\
\bottomrule
\end{tabular}
\end{center}
\vspace{-2ex}
\end{table}
\noindent Both bets pay $0$, $50$, and $100$ with equal probability. In two of the three cards, $L$ pays more than $C$; the remaining card, which delivers the greatest outcome contrast, favors $C$. Any strict preference between the two therefore reflects sensitivity to correlation itself, which has been identified in recent experiments using such same-marginal-lottery (SML) choices \citep{diecidue2020probability,bohren2024cognitive,loewenfeld2025salience}.

These observations raise the question at the center of this paper: how informative is indifference between identically distributed alternatives? We show that, under continuity, completeness, and betweenness, indifference in every same-marginal comparison at a single lottery rules out correlation sensitivity throughout the lottery domain over its support, limiting the correlation effects central to regret theory (RT, \citet{bell1982regret,loomes1982regret,loomes1987some}) and salience theory (ST, \citet{bordalo2012salience}).

Motivated by \citeauthor{fishburn1988nonlin}'s \citeyearpar{fishburn1988nonlin} comparison of bets tied to a common coin versus independent coins, we take joint outcome distributions as primitive. A correlation preference is a set $\Pi\subseteq\Delta(X\times X)$, where $X$ is the outcome set. We begin with a finite $X$ and later extend the analysis to finitely supported lotteries on general metrizable outcome spaces. Each joint distribution $\pi$, with row marginal $p$ and column marginal $q$, specifies how the two alternatives' outcomes are jointly realized. Membership $\pi\in\Pi$ means that the decision maker weakly prefers $p$ to $q$ under that joint distribution. This formulation allows incomplete or nontransitive binary choice, nests classical models in which only the marginal lotteries matter, and accommodates correlation-sensitive comparative kernels such as those in \citet{lanzani2022correlation}.

Our first result is a local-to-global dichotomy for correlation sensitivity. On the relative interior $\Delta$ of the lottery simplex, continuity, completeness, and betweenness imply exactly one of two alternatives. Either the preference is SML correlation sensitive at every $p\in\Delta$, or it is correlation insensitive at every $p\in\Delta$. 
Thus, on a finite outcome space, the three axioms impose a striking restriction on statewise comparison effects. If the decision maker is indifferent between two copies of \textit{one} full-support lottery under every possible coupling, then the pairing of received and forgone outcomes is irrelevant to choice throughout the domain. Such indifference is innocuous in the classical domain of marginal lotteries, where correlation insensitivity is implicit and reflexivity makes the decision maker indifferent between two copies of the same lottery. However, our result shows that requiring this indifference under every coupling at just one full-support lottery rules out the correlation effect which is central to RT and ST. 

Our second theorem focuses on the correlation-sensitive branch of the dichotomy. In the correlation-sensitive case, continuity, completeness, and betweenness are necessary and sufficient for a nondegenerate correlation weighted utility (CWU) representation on the entire joint-distribution simplex, including its boundary:
\[  \pi\in\Pi\iff U(\pi)=\underbrace{\mathbb E_{\pi}[\phi(x,y)]}_{\text{correlation sensitive}}+\underbrace{\psi(p,q)}_{\substack{\text{correlation insensitive}}}\geq 0\]
where $\phi$ and $\psi$ are both skew-symmetric and $\psi$ is bilinear.\footnote{Bilinearity requires $\psi(\alpha p+\beta q,r)=\alpha\psi(p,r)+\beta\psi(q,r)$ and $\psi(r,\alpha p+\beta q)=\alpha\psi(r,p)+\beta\psi(r,q)$.} The kernel $\phi$ captures sensitivity to how received and forgone outcomes are paired, while $\psi$ allows departures from classical independence that depend only on the marginal lotteries. For RT and ST, the sensitive kernel $\phi$ is the object that records, respectively, counterfactual comparison and contrast-based attention. To characterize the correlation-insensitive branch, we additionally impose projective independence and obtain the skew-symmetric bilinear (SSB) representation of \citet{fishburn1982nontransitive}. 

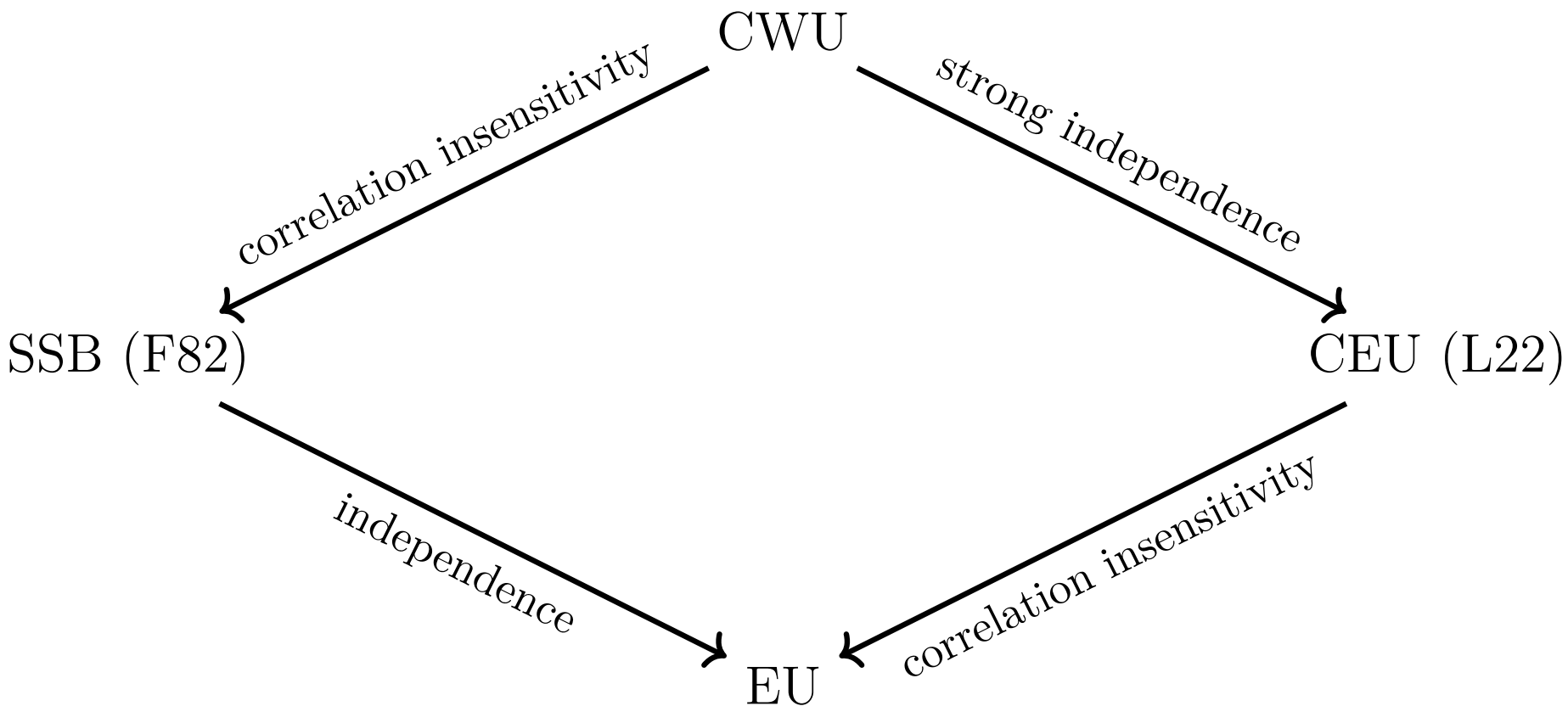
\begin{figure}[h]
\centering
\begin{tikzpicture}[scale=0.85, every node/.style={font=\large}]

\node (CWU) at (0,3) {\normalsize CWU};
\node (SSB) at (-6,0) {\normalsize SSB (F82)};
\node (CEU) at (6,0) {\normalsize CEU (L22)};
\node (EU)  at (0,-3) {\normalsize EU};

\draw[->, very thick] (CWU) -- (SSB)
    node[midway, above, sloped, font=\footnotesize] {correlation insensitivity};

\draw[->, very thick] (CWU) -- (CEU)
    node[midway, above, sloped, font=\footnotesize] {strong independence};

\draw[->, very thick] (SSB) -- (EU)
    node[midway, below, sloped, font=\footnotesize] {independence};

\draw[->, very thick] (CEU) -- (EU)
    node[midway, below, sloped, font=\footnotesize] {correlation insensitivity};

\end{tikzpicture}
\caption{CWU, SSB, CEU, and EU.}
\label{fig:routes-eu}
\end{figure}

Together, the two results combine into a uniform CWU characterization under the four axioms. The framework organizes familiar models along two dimensions: the strength of independence and the role of correlation. Strengthening betweenness to the strong independence axiom used by \citet{lanzani2022correlation}, while allowing correlation sensitivity, yields his correlation expected utility (CEU) representation, $U(\pi)=\mathbb E_{\pi}[\phi(x,y)]$.\footnote{We use the term CEU for Lanzani's correlation-sensitive representation as we place it in a broader family of correlation preferences in \citet{chenChewZhang2026correlation} which studies the general correlation preference framework.} Retaining betweenness and projective independence while imposing correlation insensitivity yields skew-symmetric bilinear (SSB) utility, $U(\pi)=\psi(p,q)$. Imposing both strong independence and correlation insensitivity yields expected utility (EU), the common specialization of CEU and SSB. Specifically, SML correlation insensitivity at just one full-support lottery reduces CEU to EU. Alternatively, adding transitivity to SSB yields the weighted utility (WU) of \citet{chew1983generalization}, and adding ordinary independence further yields EU. Figure \ref{fig:routes-eu} summarizes these relationships.

Our framework establishes a testable link between two strands of experimental research on choice under risk. The first examines preferences between identically distributed alternatives, from early tests of regret theory and similarity judgments \citep{loomes1992preferences,leland1998similarity} to recent work on probability dominance, cognitive representations, and salience \citep{diecidue2020probability,bohren2024cognitive,loewenfeld2025salience}.\footnote{Related psychological evidence examines the role of statewise conflict \citep{diederich1999conflict} and learning from experience \citep{erev2023contradictory}.} The second examines the role of correlation in Allais behavior \citep{starmer1992testing,bordalo2012salience,frydman2018role,bruhin2022risk}.\footnote{See also \citet{ostermair2022experimental,humphrey2023accepts} on Allais behavior under correlated presentations, and \citet{loewenfeld2024event} on distinguishing correlation effects from event-splitting effects.} 

Our dichotomy imposes a common restriction across these settings: a decision maker who is indifferent in every SML comparison at one full-support lottery cannot exhibit correlation-dependent Allais behavior on the same outcome support. On a common three-outcome support, the framework also yields a directional prediction. Section \ref{subsec:xsod} shows that the SML preference predicts how the two diagnostic comparisons in the HILO design of \citet{chew1986empirical} respond as dependence moves from maximal positive dependence toward independence. A preference for $C$ over $L$ predicts weakly higher probabilities choosing each of the two options associated with standard fanning out. The opposite preference for $L$, consistent with probability dominance \citep{diecidue2020probability}, predicts weakly lower probabilities; indifference predicts no pure-correlation movement. Section \ref{subsec:joint-test} develops a within-subject design that tests these linked predictions and distinguishes evidence of correlation sensitivity from violations of independence that persist under maximal positive dependence. Together, the two tasks provide complementary tests of the restrictions separating CWU and its CEU, SSB, and EU specializations.


The proof proceeds in four instructive layers. First, we fix a full-support lottery and study all comparisons involving it. Betweenness gives these comparisons a linear evaluation, and optimal transport (OT) duality identifies which parts of that evaluation can respond to changes in pairing. This shows that indifference under every coupling of two copies of the fixed lottery eliminates pairing effects from all comparisons involving that lottery (Section \ref{subsec:local-ot}). Second, we connect the evaluations obtained by fixing different lotteries. Wherever correlation matters, completeness links the two evaluations obtained by reversing the alternatives’ roles. Constrained OT duality translates this agreement into relative numerical scales, which we show are consistent across comparisons. Putting the evaluations on a common scale yields CWU: a shared term capturing pairing effects and a bilinear term depending only on the marginal lotteries (Section \ref{subsec:cycles}). Third, this representation rules out the coexistence of sensitive and insensitive full-support lotteries: comparisons near the boundary between the two regions would contradict completeness and continuity. Indifference under every coupling of two copies of \textit{one} full-support lottery therefore eliminates pairing effects throughout the interior. In the sensitive branch, we extend the representation to comparisons involving lotteries without full support (Section \ref{subsec:boundary}). Finally, in the correlation-insensitive branch, projective independence \citep{chew1994projective} supplies the additional consistency needed to obtain SSB utility. The two branches thus reveal how correlation sensitivity and projective independence provide alternative foundations for the common scale underlying CWU (Section \ref{subsec:ssb-wu}).

This paper proceeds as follows. Section \ref{section:setup} defines correlation preference, states the main results with CEU and EU as benchmark specializations, and develops the empirical predictions. Section \ref{section:proof} proves the main theorems
. Section \ref{section:experiments} reviews the experimental evidence and develops a joint SML--HILO test. Section \ref{section:discussion} extends the local-to-global argument to a more general counterfactual menu-choice model \citep{koch2026counterfactual} whose binary specialization is CEU. It also examines a correlation-sensitive analogue of dual utility, before relating the proof methods to applications of OT and graph-theoretic integration. Section \ref{section:conclusion} concludes. Appendix \ref{sec:finite-support-extension} extends the main results to finitely supported distributions on a general metrizable outcome space.

\section{Correlation Preference and Main Results}\label{section:setup}

For simplicity, Sections \ref{section:setup}--\ref{section:proof} formulate the baseline model for a finite real outcome set $X$ with $|X|=n$. Section \ref{sec:finite-support-extension} gives the corresponding extension to a general metrizable outcome space with finitely supported probability measures used by \citet{lanzani2022correlation}.

\subsection{Correlation Preference and Correlation (In-)Sensitivity}\label{subsec:domain}

In binary choice under risk, it is customary to denote a (nonstrict) preference of the decision maker (DM) for a lottery $p$ over another lottery $q$ using the notation $p\succeq q$. In this regard, the domain of choice is implicitly $\Delta X\times\Delta X$, where $\Delta X$ denotes the set of lotteries on $X$. The preference under risk is then described by a binary relation $R\subseteq\Delta X\times\Delta X$, called a \textit{preference set}, with $(p,q)\in R$ meaning that DM weakly prefers $p$ to $q$.
\begin{table}[h]
\centering
\begin{tabular}{|c|c|c|c|} 
 \hline 
 $\pi$ & $y_1$ & $\cdots$  & $y_l$ \\ \hline
 $x_1$ & $\pi(x_1,y_1)$ & $\cdots$ & $\pi(x_1,y_l)$ \\ \hline
 $\vdots$ & $\vdots$ & $\vdots$ & $\vdots$ \\ \hline
 $x_m$ & $\pi(x_m,y_1)$& $\cdots$ & $\pi(x_m,y_l)$ \\\hline
\end{tabular}
\vspace{-1ex}
\end{table}

To accommodate the possibility of binary preference being sensitive to correlation between lotteries, we expand the domain from $\Delta X\times\Delta X$ to $\Delta(X\times X)$, comprising joint densities defined on $X\times X$. A joint density $\pi$ illustrated in the tabular form (with possibly different supports of marginals) can be written as $\sum_{x,y\in X}\pi(x,y)\delta(x,y)$, where $\delta(x,y)$ denotes point mass at $(x,y)$: receiving $x$ against receiving $y$ with certainty. Let $p(x)=\sum_{y\in X}\pi(x,y)$ and $q(y)=\sum_{x\in X}\pi(x,y)$ denote the \textit{row} and \textit{column} marginals, respectively. For $p,q\in\Delta X$, let
\[
 \Gamma(p,q):=\{\pi\in\Delta(X\times X):(\pi_1,\pi_2)=(p,q)\}.
\]

Write $p\succeq^\pi q$ when DM weakly $p$ over $q$ \textit{under joint density} $\pi$. In binary choice between row and column outcomes, changing the coupling $\pi$ within $\Gamma(p,q)$ affects how the received and forgone outcomes are jointly realized while preserving both marginal lotteries. In line with the practice of identifying $\succeq$ with its preference set $R \subset
\Delta X\times\Delta X$, it is convenient to associate $\succeq^\pi$ with the corresponding preference set $\Pi \subset \Delta(X \times X)$. A \emph{correlation preference} is thus represented by a set $\Pi\subseteq\Delta(X\times X)$. Membership $\pi\in\Pi$ means that DM weakly prefers receiving its row marginal $\pi_1=\sum_{x\in X}p(x)\delta_x$ to receiving its column marginal $\pi_2=\sum_{y\in X}q(y)\delta_y$, when the outcomes are jointly distributed according to $\pi$. 

To describe the induced strict preference, let $\pi^{\text{\tiny{T}}}$ denote the transpose of $\pi$ indicating the switch in the roles of row and column marginals. Write $\hat{\Pi}=\{\pi:\pi \in \Pi,\ \pi^{\text{\tiny{T}}}\notin \Pi \}$ as the \textit{strict preference} set, $\tilde{\Pi}=\Pi \setminus \hat{\Pi}$ the \textit{indifference set}, and $\check{\Pi}$ the complement of $\Pi$ in $\Delta(X\times X)$. For instance, the preference is complete if $\pi \in \check\Pi$ implies $\pi^{\text{\tiny{T}}}\in\Pi$. Notice the distinct role of \textit{indifference set} $\tilde{\Pi}$ from that of \textit{indifference curves} in the two settings: under transitivity and reflexivity, indifference curves \textit{partition} the entire domain, while in the correlation setting, points in $\tilde{\Pi}$ do not entertain such an equivalence relation. The table below summarizes the relation between two sets of notations.

\begin{table}[h]
    \centering
    \begin{tabular}{ccccc} \hline
                  $p$ vs. $q$ & $p\succeq^\pi q$ & $p\succ^\pi q$ & $p\sim^\pi q$ &$p\nsucceq^\pi q$\\ \hline
                  $\pi\in\cdot$& $\Pi$ & $\hat\Pi$ & $\tilde\Pi$ & $\check\Pi$\\ \hline 
    \end{tabular}
\vspace{-1ex}
\end{table}
We now introduce the \textit{induced correlation-insensitive} comparisons from $\Pi$. Write $p\succeq^*q$ if $\Gamma(p,q)\subseteq\Pi$, and $p\sim^*q$, $p\succ^*q$, and $p\nsucceq^*q$ likewise. Note that this induced preference on $\Delta X$ need not be complete.

\begin{defi}[Correlation Insensitivity]\label{def:ci}
The preference $\Pi$ is correlation insensitive (CI) for $(p,q)$ if membership in $\Pi$ is constant on $\Gamma(p,q)$: either $\Gamma(p,q)\subseteq\Pi$ or $\Gamma(p,q)\cap\Pi=\varnothing$. It is correlation insensitive at $p$ if this holds for every $q\in\Delta X$, and globally correlation insensitive on a set $P\subseteq\Delta X$ if it is correlation insensitive at every $p\in P$.
\end{defi}
The preference is correlation sensitive (CS) at $p$ if it is not CI at $p$. The following local version of correlation insensitivity plays the key role in this paper.

\begin{defi}[SML Correlation Insensitivity]\label{def:smlci}
The preference is same-marginal-lottery correlation insensitive (SML-CI) at $p$ if $p\sim^* p$, i.e., every element of $\Gamma(p,p)$ is indifferent. 
\end{defi}
Let $\Delta$ be the relative interior of $\Delta X$ containing all full-support marginals. The following partition is instrumental to our analysis:
\[
 \Delta^I:=\{p\in\Delta:p\text{ is SML-CI}\},\qquad
 \Delta^S:=\Delta\setminus\Delta^I.
\]

\subsection{Behavioral Axioms}\label{subsec:axioms}

We maintain continuity, completeness, and betweenness for the local-to-global theorem.

\begin{axm}[Continuity]\label{axm:continuity}
$\Pi$ is closed relative to $\Delta(X\times X)$.
\end{axm}

Equivalently, a limit of weakly accepted comparisons remains weakly accepted. This is the usual topological regularity condition, now imposed on joint distributions rather than on pairs of marginal lotteries.

\begin{axm}[Completeness]\label{axm:completeness}
For every $\pi\in\Delta(X\times X)$, $\pi\in\check\Pi$ implies $\pi^{\text{\tiny{T}}}\in\Pi$.
\end{axm}

Like its classical counterpart, completeness requires every binary comparison to be resolved in at least one direction. Transposition simply exchanges the received and forgone outcomes. 

\begin{axm}[Betweenness]\label{axm:betweenness}
For $\pi,\pi'\in\Delta(X\times X)$ with $\pi_1=\pi'_1$ and every $\alpha\in(0,1)$:
\begin{enumerate}[noitemsep,topsep=2pt]
 \item $\pi,\pi'\in\widetilde\Pi$ implies $\alpha\pi+(1-\alpha)\pi'\in\widetilde\Pi$;
 \item $\pi\in\widehat\Pi$ and $\pi'\in\Pi$ implies $\alpha\pi+(1-\alpha)\pi'\in\widehat\Pi$;
 \item $\pi\in\check\Pi\cup\widetilde\Pi$ and $\pi'\in\check\Pi$ implies $\alpha\pi+(1-\alpha)\pi'\in\check\Pi$.
\end{enumerate}
\end{axm}

In the $\succ^\pi$ notation, the second clause says $p\succ^\pi q$ and $p\succeq^{\pi'}q'$ imply $p\succ^{\pi_\alpha}q_\alpha$, where $\pi_\alpha:=\alpha\pi+(1-\alpha)\pi'$ and likewise for $q_\alpha$. The first marginal remains $p$; the first and third clauses have the analogous interpretations. Geometrically, the better-than and worse-than regions are convex within each fixed-first-marginal fiber. Unlike strong independence, betweenness allows the local evaluation to vary with the first marginal. In RT and ST terms, it allows the comparative force of a given outcome pair dependent on the lottery at stake.


\subsection{Main Theorems and Benchmark Preview}\label{subsec:main-results}

\begin{thm}[Local-to-Global Correlation Dichotomy]\label{thm:dichotomy}
If the correlation preference satisfies continuity, completeness, and betweenness, then exactly one of the following holds:
\begin{enumerate}[noitemsep,topsep=2pt]
 \item $\Delta^I=\Delta$. The preference is correlation insensitive at every $p\in\Delta$.
 \item $\Delta^S=\Delta$.
\end{enumerate}
\end{thm}
The theorem is stronger than the pointwise equivalence between SML-CI at $p$ and correlation insensitivity at $p$. Proposition \ref{prop:pointwise-ci} below isolates that preliminary implication: under the maintained continuity and completeness conditions, betweenness alone implies that SML-CI at a full-support $p\in\Delta$ leads to CI at every comparison involving $p$. Thus, the previous $\Delta^I\bigcup\Delta^S$ partition has implications beyond the SML choices in the global preference domain. Unlike the similar CI-to-EU properties obtained by \citet{lanzani2022correlation}, neither independence nor any other cross-fiber independence condition is needed for this fixed-marginal step. Theorem \ref{thm:dichotomy} now adds the genuinely global conclusion that one locally insensitive interior marginal cannot coexist with a sensitive region elsewhere.

For matrices $A$ and $\pi$, write $\langle A,\pi\rangle:=\sum_{x,y\in X}A(x,y)\pi(x,y)$. 
\begin{defi}[Correlation Weighted Utility]\label{def:cwu}
The preference admits a correlation weighted utility (CWU) representation if there are skew-symmetric matrices $\phi$ and $\psi$ such that $\pi\in\Pi\iff U(\pi):=\langle\phi,\pi\rangle+\pi_1^{\text{\tiny{T}}}\psi\pi_2\geq0$.
\end{defi}
Equivalently, $\phi$ is a skew-symmetric kernel on $X\times X$ and, with the customary abuse of notation, $\psi(p,q):=p^{\text{\tiny{T}}}\psi q$ is a skew-symmetric bilinear functional of the marginals. Let $\Sspace:=\{u\one^{\text{\tiny{T}}}-\one v^{\text{\tiny{T}}}:u,v\in\R^n\}$ be the space of additively separable matrices, where $\mathbb 1$ is the vectors of all ones. A comparative kernel $\phi$ is \emph{nondegenerate} if $\phi\notin\Sspace$; a representation with correlation kernel $\phi$ is nondegenerate when its kernel is. Thus nondegeneracy means that the correlation term cannot be absorbed into a function of the two marginals. A skew-symmetric member of $\Sspace$ has the expected-utility-difference form $u\one^{\text{\tiny{T}}}-\one u^{\text{\tiny{T}}}$ and can be absorbed into $\psi$. Thus a \emph{degenerate CWU} can be normalized to $\phi=0$, which gives precisely SSB utility. Setting $\psi=0$ instead gives \emph{CEU}.

\begin{thm}[Correlation-Sensitive CWU]\label{thm:cwu}
A correlation preference satisfies continuity, completeness, and betweenness, and is correlation sensitive at some $p\in\Delta$ if and only if it admits a nondegenerate CWU representation on $\Delta(X\times X)$. In this case, $\Delta^S=\Delta$.
\end{thm}

Theorem \ref{thm:cwu} isolates the substantive correlation-sensitive conclusion: the three maintained axioms already yield one nonseparable kernel on the entire domain. Projective independence will be introduced only when the CI branch is represented in Subsection \ref{subsec:ssb-wu}. Existing models then arise by restricting the two possible branches of CWU. Strong independence yields CEU below, which includes the relevant binary ST and RT specifications of \citet{lanzani2022correlation}. Within CEU,  SML-CI at one full-support lottery reduces CEU to EU. The insensitive branch leads from SSB to WU and EU as developed in Subsection \ref{subsec:ssb-wu}.

\subsection{CEU and EU as Benchmark Specializations}\label{subsec:ceu}

We first discuss the special case of correlation expected utility, introduced and axiomatized by \citet{lanzani2022correlation}, as the linear specialization of CWU. Its comparative kernel encompasses the state-by-state forces central to RT and ST while remaining linear in the joint distribution.

\begin{defi}[Correlation Expected Utility]\label{def:ceu}
The preference admits a correlation expected utility (CEU) representation if there is a skew-symmetric matrix $\phi$ such that $\pi\in\Pi\iff\langle\phi,\pi\rangle\geq0$, with equality if and only if $\pi$ is indifferent.
\end{defi}

The following yields a sharp restriction within CEU and serves as an illustrative example for the betweenness results. 

\begin{cor}[SML-CI under CEU]\label{cor:ceu-eu}
If a CEU preference is SML-CI at some $p\in\Delta$, then it admits an EU representation.
\end{cor}
Thus a complete CEU preference cannot be both genuinely correlation sensitive and SML-CI at one full-support lottery. In the terminology fixed above, every non-EU CEU kernel is nondegenerate and therefore belongs to the sensitive branch of Theorem \ref{thm:dichotomy}. Neither transitivity nor correlation insensitivity on the entire domain is required, contrasting a similar CI-to-EU result observed by \citet{lanzani2022correlation}.\footnote{\citet{lanzani2022correlation} points out that a CEU model ``\textit{coincides}'' with SSB ``\textit{when restricted to the comparison between independent lotteries}'', and writes, ``\textit{Theorem 1 provides an alternative set of axioms for the SSB model}''. From the perspective of shrinking the domain to the set of independent joint densities, strong independence, if defined on joint distributions as it current version, is not applicable, since a convex combination of two arbitrary \textit{independent} joint densities may not be independent. Instead, one would need to impose correlation insensitivity so that the ``restriction'' is well defined. In this regard, our dichotomy theorem and the corollary clarify that a correlation-insensitive CEU reduces to EU, not SSB utility.} We further discuss a generalization of CEU in Section \ref{subsec:kis-context-projection} based on \citet{koch2026counterfactual}, which also bears such bang-bang correlation sensitivity without transitivity.

\subsection{Outcome-Difference Sensitivity and Correlated-Allais Predictions}\label{subsec:xsod}

As an important implication of betweenness and the above theorems, we formulate the outcome-difference sensitivity classification of \citet{loewenfeld2025salience} directly in terms of SML comparisons.\footnote{We discuss details on experimental design, calculations and predictions, and existing evidence in Section~\ref{section:experiments}. } Fix three outcomes $x_L<x_M<x_H$, let $p_{\textit{\tiny LMH}}:=\frac13(\delta_{x_L}+\delta_{x_M}+\delta_{x_H})$, and define the oriented SML cycle
\begin{equation}
 \pi_{\textit{\tiny LMH}}:=\frac13\bigl(\delta_{(x_L,x_M)}+\delta_{(x_M,x_H)}+\delta_{(x_H,x_L)}\bigr)
 \in\Gamma(p_{\textit{\tiny LMH}},p_{\textit{\tiny LMH}}).
 \label{eq:sml-cycle}
\end{equation}
The row alternative loses by the two adjacent outcome differences and wins by the largest outcome difference. Its transpose reverses these comparisons.

\begin{defi}[Local and Global xSOD]\label{def:xsod}
For the fixed support $\{x_L,x_M,x_H\}$, the preference satisfies
\begin{enumerate}[label=\textup{(\roman*)},noitemsep,topsep=3pt]
 \item local increasing sensitivity to outcome differences (local ISOD) if $\pi_{\textit{\tiny LMH}}\in\widehat\Pi$;
 \item local constant sensitivity to outcome differences (local CSOD) if $\pi_{\textit{\tiny LMH}}\in\widetilde\Pi$;
 \item local decreasing sensitivity to outcome differences (local DSOD) if $\pi_{\textit{\tiny LMH}}\in\check\Pi$.
\end{enumerate}
For \emph{x}$\in\{I,C,D\}$, the preference is \emph{xSOD} if it is local \emph{xSOD} on every ordered three-point support $x_L<x_M<x_H$ in $X$.
\end{defi}

Thus, the $C$ (contrast) bet choice in the introduction corresponds to local ISOD, while $L$ (likelihood) bet choice and indifference correspond to local DSOD and CSOD respectively. Under CWU, write
\begin{equation}
 \Delta_\phi(x_L,x_M,x_H)
 :=\phi(x_H,x_L)-\phi(x_H,x_M)-\phi(x_M,x_L).
 \label{eq:delta-phi}
\end{equation}
Because $\psi(p_{\textit{\tiny LMH}},p_{\textit{\tiny LMH}})=0$, equation \eqref{eq:sml-cycle} has value $U(\pi_{\textit{\tiny LMH}})=\Delta_\phi/3$. Hence local ISOD, CSOD, and DSOD are equivalent, respectively, to $\Delta_\phi>0$, $\Delta_\phi=0$, and $\Delta_\phi<0$. This recovers the classification of \citet{loewenfeld2025salience} without requiring their maintained independence axiom and CEU framework.

The same index yields a correlated-Allais prediction using the HILO design of \citet{chew1986empirical}. Starting from the original comparison $O$, the $I$-, $L$-, and $H$-situations mix each alternative with a common consequence $x_M$, $x_L$, and $x_H$, respectively. In Table \ref{tab:hilo-marginals}, $\alpha\in(0,1)$ is the probability of retaining the original lottery, $1-\alpha$ is the common-consequence probability, and $\beta\in(0,1)$ is the high-outcome probability in $B_O$; $[x]$ stands for a sure outcome at $x$. The four choices retain the original sixteen-pattern HILO classification; in particular, $A_OA_IB_LA_H$ is the standard fanning-out pattern.

\begin{table}[h]
\centering
\caption{Marginal lotteries in the HILO block}
\label{tab:hilo-marginals}
\small
\begin{tabular}{@{}cll@{}}
\toprule
Situation & Alternative $A$ & Alternative $B$ \\
\midrule
$O$ & $[x_M]$ & $\beta[x_H]+(1-\beta)[x_L]$ \\
$I$ & $[x_M]$ & $\alpha\beta[x_H]+(1-\alpha)[x_M]+\alpha(1-\beta)[x_L]$ \\
$L$ & $\alpha[x_M]+(1-\alpha)[x_L]$ & $\alpha\beta[x_H]+(1-\alpha\beta)[x_L]$ \\
$H$ & $\alpha[x_M]+(1-\alpha)[x_H]$ & $(1-\alpha+\alpha\beta)[x_H]+\alpha(1-\beta)[x_L]$ \\
\bottomrule
\end{tabular}
\end{table}
To test the correlation preference, parameterize the coupling by $\lambda\in[0,1]$, with $\lambda=1$ denoting maximal positive dependence and $\lambda=0$ independence. Let $\pi_z^{\mathrm{ind}}$ be the product coupling, $\pi_z^+$ the comonotone coupling, and define $ \pi_z^\lambda=(1-\lambda)\pi_z^{\mathrm{ind}}+\lambda\pi_z^+$ for $\lambda\in\{0,\tfrac12,1\}$. 
Holding the Allais marginals fixed, moving from maximal dependence toward independence changes the two diagnostic HILO utilities by
\begin{align}
 U_L^\lambda-U_L^1
 & =-(1-\lambda)\alpha\beta(1-\alpha)\Delta_\phi,\label{eq:xsod-allais-L}\\
 U_H^\lambda-U_H^1
 & =(1-\lambda)\alpha(1-\beta)(1-\alpha)\Delta_\phi.\label{eq:xsod-allais-H}
\end{align}
It is natural to assume that for subject $i$ and situation $z$, $\Pr_i(A\mid z,\lambda)=F_{iz}(U_i(\pi_z^\lambda))$, where $F_{iz}$ is nondecreasing and does not vary with $\lambda$. Under this condition, the probabilities of the two constituent standard-direction choices, and hence their average, are weakly increasing, constant, or weakly decreasing, respectively, as one moves from maximal positive dependence toward independence; the movements are strict when $F_{iz}$ is strictly increasing over the relevant range and $\Delta_\phi\neq0$. Thus ISOD moves both choices in the standard, light Allais direction as correlation falls; CSOD produces no pure-correlation change; and DSOD moves both choices in the reverse, heavy direction. 
Section \ref{subsec:joint-test} develops the joint experimental test in further detail. 
\begin{prop}[SML--Allais Prediction]\label{prop:sml-allais}
Consider the oriented SML and HILO comparisons above. For a CWU preference DM with $F_z$ above, the probabilities of the fanning-out choices $(B_L,A_H)$ are both weakly decreasing (resp. increasing, constant) in the level of correlation $\lambda$ if the decision maker exhibits ISOD (resp. DSOD, CSOD) in the SML choice.
\end{prop}

\section{Proof Architecture and Global Representation}\label{section:proof}

The proof proceeds in four stages. Subsection \ref{subsec:local-ot} uses fiberwise separation to obtain a CBU representation and then applies constrained OT duality to identify SML-CI with additive separability of the local kernel, and hence with one-marginal CI. Subsection \ref{subsec:cycles} supplies the main technical integration step: the \textit{sensitivity graph} links nonseparable local kernels, completeness determines positive edge scales, and a product-one condition around every cycle in the graph generates a cardinal potential. After this normalization, local skew identities patch into the component-wise CWU representation summarized in Proposition \ref{prop:component-cwu}. Subsection \ref{subsec:boundary} uses that representation to rule out coexistence of CS and CI components and proves Theorems \ref{thm:dichotomy} and \ref{thm:cwu}. Subsection \ref{subsec:ssb-wu} then treats the insensitive branch as one sequence: projective independence normalizes it to SSB, CI-transitivity reduces SSB to WU, and Proposition \ref{prop:cwu-uniform} closes with the uniform representation.

\subsection{Local Linear Structure and Optimal-Transport Separability}\label{subsec:local-ot}

For $p\in\Delta X$, define the fixed-first-marginal fiber $\Delta_p:=\{\pi\in\Delta(X\times X):\pi_1=p\}$. On this fiber, betweenness makes the strict, indifferent, and rejected regions convex. When both strict regions are nonempty, their common boundary is the indifference set; in the one-sided cases, the indifference set is an exposed face of the fiber. Finite-dimensional separation therefore supplies a local kernel $\phi_p$. 
This gives the intermediate representation used in proving both Theorems \ref{thm:dichotomy} and \ref{thm:cwu}.

\begin{defi}[Correlation Betweenness Utility]\label{def:cbu}
The preference admits a correlation betweenness utility (CBU) representation if, for every $p\in\Delta X$, there is a matrix $\phi_p$ such that for every $\pi\in\Delta_p$, $\pi\in\Pi\iff \langle\phi_p,\pi\rangle\geq0$, and $\pi\in\widetilde\Pi\iff \langle\phi_p,\pi\rangle=0$.
\end{defi}

\begin{prop}[Fiberwise Linear Representation]\label{prop:cbu}
A continuous and complete correlation preference that satisfies betweenness admits a CBU representation. Conversely, every CBU representation satisfies betweenness.
\end{prop}
Appendix \ref{appendix:fiber} gives the proof. 

The main conclusion of this subsection is the following pointwise equivalence. It turns the local SML-CI observation at $p$ into CI for every comparison having first marginal $p$.

\begin{prop}[Pointwise CI Equivalence under CBU]\label{prop:pointwise-ci}
Let $p\in\Delta$. For a complete  CBU preference, the following are equivalent:
\begin{enumerate}[noitemsep,topsep=2pt]
 \item $p\in\Delta^I$;
 \item the preference is correlation insensitive at $p$;
 \item $\phi_p\in\Sspace$.
\end{enumerate}
\end{prop}

Define the zero-marginal tangent space $\Tspace:=\{Z\in\R^{n\times n}:Z\one=0,\ Z^{\text{\tiny{T}}}\one=0\}$. Proposition \ref{prop:pointwise-ci} rests on the following technical lemma: perturbing a full-support coupling in every direction in $\Tspace$ forces the evaluating kernel into $\Tspace^\perp=\Sspace$. It is used repeatedly in the pointwise and subsequent local-to-global arguments.

\begin{lem}[Transport Separability]\label{lem:transport-separable}
Let $p,q\in\Delta$ and $\phi\in\R^{n\times n}$. If $\langle\phi,\pi\rangle=0$ for every $\pi\in\Gamma(p,q)$, then $\phi\in\Sspace$; equivalently, $\phi(x,y)=u(x)-v(y)$ for some $u,v:X\to\R$.
\end{lem}

\begin{proof}
Consider the finite OT problem $\max_{\pi\in\Gamma(p,q)}\langle\phi,\pi\rangle$. Its value is zero by assumption, and the independent joint distribution $\pi^0:=pq^{\text{\tiny{T}}}$ is a full-support optimum. Writing the row and column constraints as $\sum_y\pi(x,y)=p(x)$ and $\sum_x\pi(x,y)=q(y)$, respectively, its Kantorovich dual is
\[
 \min_{u,v:X\to\R}\{u\cdot p-v\cdot q:u(x)-v(y)\geq\phi(x,y)\text{ for all }x,y\in X\}.
\]
Strong duality gives an optimal pair $(u,v)$ with value zero. Let $s(x,y):=u(x)-v(y)-\phi(x,y)\geq0$ denote the dual slack. Feasibility of $\pi^0$ and equality of the primal and dual values give $0=u\cdot p-v\cdot q-\langle\phi,\pi^0\rangle=\sum_{x,y\in X}p(x)q(y)s(x,y)$. Since $p$ and $q$ have full support, every coefficient $p(x)q(y)$ is strictly positive. Hence $s(x,y)=0$ for every $x,y\in X$, proving $\phi(x,y)=u(x)-v(y)$ throughout $X\times X$.
\end{proof}

The argument further implies $\Tspace^\perp=\Sspace$. Indeed, $\Sspace\subseteq\Tspace^\perp$ is immediate. Conversely, if $B\perp\Tspace$, then $\langle B,\pi\rangle$ is constant on $\Gamma(p,q)$ for any fixed $p,q\in\Delta$; subtracting this constant from every entry of $B$ reduces the claim to Lemma \ref{lem:transport-separable}. If $B$ is skew-symmetric, additive separability and the zero diagonal give $B(x,y)=u(x)-u(y)$.

\begin{proof}[Proof of Proposition \ref{prop:pointwise-ci}]
If $p\in\Delta^I$,  CBU gives $\langle\phi_p,\pi\rangle=0$ throughout $\Gamma(p,p)$. Lemma \ref{lem:transport-separable} yields $\phi_p=u\one^{\text{\tiny{T}}}-\one v^{\text{\tiny{T}}}$. Hence, for every $q$ and $\pi\in\Gamma(p,q)$, $\langle\phi_p,\pi\rangle=u\cdot p-v\cdot q$, which is independent of the correlation. Thus the preference is CI at $p$.

Conversely, if the preference is CI at $p$, all elements of $\Gamma(p,p)$ have the same membership status. The symmetric distribution $pp^{\text{\tiny{T}}}$ is indifferent by Lemma \ref{lem:transpose}. Completeness therefore implies that every element of $\Gamma(p,p)$ is indifferent, so $p\in\Delta^I$. The equivalence with separability follows.
\end{proof}

Additive separability is thus exactly the local component that cannot detect changes in correlation. Proposition \ref{prop:pointwise-ci} hence links the partition $\Delta=\Delta^I\cup\Delta^S$ to the separability of the local kernels.

\subsection{Sensitivity Cycles and the Component-wise CWU Representation}\label{subsec:cycles}

\begin{defi}[Sensitivity Graph]\label{def:sensitive-pair}
For $p,q\in\Delta^S$, write $(p,q)\in E$ if $ \Gamma(p,q)\cap\widehat\Pi\neq\varnothing$ and $\Gamma(p,q)\cap\check\Pi\neq\varnothing$. The graph with node set $\Delta^S$ and edge set $E$ is the sensitivity graph.
\end{defi}
Lemma \ref{lem:transpose} in Appendix \ref{appendix:ceu} shows that $E$ must be symmetric. More importantly, the sensitivity graph is locally complete around every sensitive marginal.

\begin{lem}[Local Sensitivity Neighborhoods]\label{lem:local-sensitivity}
For every $p\in\Delta^S$, there is a relatively open neighborhood $N_p\subseteq\Delta^S$ such that $(r,s)\in E$ for every $r,s\in N_p$. Consequently, $\Delta^S$ is open, and each of its topological connected components is connected in the sensitivity graph by a finite edge path.
\end{lem}

The proof is given in Appendix \ref{appendix:local-sensitivity}. The sensitivity graph and the next lemma are the main technical devices in this subsection. Completeness first determines the relative cardinal scale of two local CBU kernels on every edge; consistency around cycles then integrates those pairwise scales into one cardinal potential.

\begin{lem}[Cardinal Normalization and Generalized Skew Symmetry]\label{lem:pairwise-scale}
On every connected component $C$ of the sensitivity graph, the local CBU kernels admit positive normalizations $H_p:=\gamma(p)\phi_p$, $p\in C$, such that for every edge $(p,q)\in E$ there are functions $u_{pq},v_{pq}:X\to\R$ satisfying $H_p(x,y)+H_q(y,x)=u_{pq}(x)-v_{pq}(y)$ for all $x,y\in X$, with $u_{pq}\cdot p-v_{pq}\cdot q=0$.
\end{lem}
Thus the edge relation takes the generalized skew-symmetry form with its pairwise coefficient normalized to one. The proof below derives the normalization from constrained OT duality and a product-one condition on sensitivity cycles.
\begin{proof}
Fix $(p,q)\in E$ and abbreviate $ L(\pi):=\langle\phi_p,\pi\rangle$, $M(\pi):=\langle\phi_q^{\text{\tiny{T}}},\pi\rangle=\langle\phi_q,\pi^{\text{\tiny{T}}}\rangle$.

We first obtain a full-support indifferent point in $\Gamma(p,q)$. Correlation sensitivity of the pair supplies a strict and a rejected joint distribution. Mixing each by a sufficiently small amount with the full-support distribution $pq^{\text{\tiny{T}}}$ preserves its status by openness and makes it full support. The segment joining the two resulting points remains full support. Since $L$ is linear and changes sign along this segment, it contains a full-support $\pi^0$ with $L(\pi^0)=0$.  CBU makes $\pi^0$ indifferent, and Lemma \ref{lem:transpose} gives $M(\pi^0)=0$.

Consider the constrained OT problem
\[
 \min_{\pi\in\Gamma(p,q)}L(\pi)
 \qquad\text{subject to }M(\pi)\leq0.
\]
Its value is zero. If a feasible $\pi$ satisfies $M(\pi)<0$, then $\pi^{\text{\tiny{T}}}$ is rejected. Completeness makes $\pi$ strict, so $L(\pi)>0$. If $M(\pi)=0$, then $\pi^{\text{\tiny{T}}}$ and hence $\pi$ are indifferent, so $L(\pi)=0$. The point $\pi^0$ attains zero.

Let $b$ be the nonnegative multiplier on the inequality constraint. The dual is
\[
 \max_{b\ge 0,u,v:X\to\R}u\cdot p-v\cdot q
 \quad\text{subject to}\quad
 u(x)-v(y)\leq\phi_p(x,y)+b\phi_q(y,x)\ \forall x,y\in X.
\]
Finite-dimensional duality applies. Choose an optimum $(b_{pq},\bar u_{pq},\bar v_{pq})$. Since the primal optimum $\pi^0$ has full support, complementary slackness makes every cell constraint bind, so $\phi_p(x,y)+b_{pq}\phi_q(y,x)=\bar u_{pq}(x)-\bar v_{pq}(y)$ for every $x,y\in X$; equality of primal and dual values gives $\bar u_{pq}\cdot p-\bar v_{pq}\cdot q=0$. The potentials are determined up to a common additive constant. Note that the multiplier is positive: if $b_{pq}=0$, the pairwise decomposition makes $\phi_p$ separable, contrary to $p\in\Delta^S$ and Proposition \ref{prop:pointwise-ci}. It is unique as well: if $b$ and $b'$ both satisfy the conclusion, then $(b-b')\phi_q^{\text{\tiny{T}}}\in\Sspace$; since $\phi_q\notin\Sspace$, one has $b=b'$.

It remains to integrate these edge scales. Let $\mathcal Q:=\R^{n\times n}/\Sspace$, write $[A]$ for the class of $A$, and let $J[A]:=[A^{\text{\tiny{T}}}]$. Transposition preserves $\Sspace$, so $J$ is a well-defined involution. The edge identity gives $[\phi_p]+b_{pq}J[\phi_q]=0$. For each $p\in\Delta^S$, the self-pair $(p,p)$ is an edge: a nonindifferent element of $\Gamma(p,p)$ and its transpose have opposite strict status. The quotient identity for this self-edge and its image under $J$ imply $b_{pp}=1$ and $J[\phi_p]=-[\phi_p]$. Hence $[\phi_p]=b_{pq}[\phi_q]$ on every edge.

For a directed sensitivity cycle $c=(p_0,p_1,\ldots,p_k=p_0)$, repeated substitution gives \([\phi_{p_0}]=\left(\prod_{r=0}^{k-1}b_{p_rp_{r+1}}\right)[\phi_{p_0}]\). Because $[\phi_{p_0}]\neq0$, the product around every cycle is one. Choose a root $p_0\in C$, set $\gamma(p_0)=1$, and define $\gamma(p)$ as the product of edge coefficients along any finite sensitivity path from $p_0$ to $p$. The cycle identity makes this path independent and gives $b_{pq}=\gamma(q)/\gamma(p)$. Thus $H_p:=\gamma(p)\phi_p$ remains a local representation and satisfies $[H_p]=[H_q]$ on every edge. Graph connectivity supplies a common nonzero class $h$ for all $p\in C$, and $J(h)=-h$. Finally, multiply the pairwise edge identity by $\gamma(p)$ and set $u_{pq}:=\gamma(p)\bar u_{pq}$ and $v_{pq}:=\gamma(p)\bar v_{pq}$. The relation $b_{pq}=\gamma(q)/\gamma(p)$ then gives $H_p(x,y)+H_q(y,x)=u_{pq}(x)-v_{pq}(y)$ and $u_{pq}\cdot p-v_{pq}\cdot q=0$, as claimed.
\end{proof}

The remaining step converts the common quotient class into one skew kernel and patches the separable residuals into a bilinear marginal term.

\begin{prop}[Component-Wise CWU]\label{prop:component-cwu}
Let $C$ be a connected component of $\Delta^S$. There exist skew-symmetric matrices $\phi_C$ and $\psi_C$, with $\phi_C\notin\Sspace$, such that, for every $p\in C$, every $q\in\Delta X$, and every $\pi\in\Gamma(p,q)$, $\pi\in\Pi\iff \langle\phi_C,\pi\rangle+p^{\text{\tiny{T}}}\psi_Cq\geq0$ with equality if and only if $\pi$ is indifferent.
\end{prop}

The proof is given in Appendix \ref{appendix:component-cwu}. Lemma \ref{lem:pairwise-scale} has already done the cardinal work. One chooses a skew-symmetric representative $\phi_C$ of the common quotient class and writes each normalized kernel as $\phi_C$ plus a separable residual. The zero-expectation edge identities make those residual expectations locally bilinear and skew-symmetric in marginals; the locally complete neighborhoods in Lemma \ref{lem:local-sensitivity} then patch them into one matrix $\psi_C$. Importantly, the resulting identity is a vector identity, so the representation applies to every second marginal $q\in\Delta X$, not merely to $q\in C$. This unrestricted scope is what makes the boundary argument below possible.

\subsection{The Boundary Argument and the Correlation-Sensitive Representation}\label{subsec:boundary}

The strengthened scope of Proposition \ref{prop:component-cwu} makes the final local-to-global step direct. This allows us to compare marginals in a sensitive component with a marginal on its relative boundary, which yields the contradiction below.

\begin{proof}[Proof of Theorem \ref{thm:dichotomy}]
If $\Delta^S=\varnothing$, then $\Delta^I=\Delta$. Proposition \ref{prop:pointwise-ci} makes the preference correlation insensitive at every $p\in\Delta$, giving the first branch. Now suppose instead that $\Delta^S\neq\varnothing$, and let $C$ be a connected component of $\Delta^S$. Proposition \ref{prop:component-cwu} supplies skew-symmetric matrices $\phi$ and $\psi$ with $\phi\notin\Sspace$, such that $U_C(\pi):=\langle\phi,\pi\rangle+\pi_1^{\text{\tiny{T}}}\psi\pi_2$ represents every comparison whose first marginal belongs to $C$, regardless of its second marginal.

We prove that $C=\Delta$. Suppose to the contrary that $C$ is proper. Since $C$ is a nonempty open subset of the connected set $\Delta$, its relative boundary in $\Delta$ is nonempty. Choose $p\in\partial_\Delta C$ and a sequence $p^m\in C$ with $p^m\to p$. The boundary point belongs to $\Delta^I$: if $p\in\Delta^S$, openness of $\Delta^S$ would give a connected neighborhood of $p$ contained in $\Delta^S$ and intersecting $C$, contrary to maximality of $C$. 

Because $\phi\notin\Sspace=\Tspace^\perp$, choose $Z\in\Tspace$ with $h:=\langle\phi,Z\rangle>0$. Since $p$ has full support and $p^m\to p$, there is a fixed $\varepsilon>0$ such that, for $m$ sufficiently large, $\pi_m^\pm:=p^mp^{\text{\tiny{T}}}\pm\varepsilon Z$ are nonnegative. The perturbation $Z$ gives $\pi_m^\pm\in\Gamma(p^m,p)$. By the component-wise representation,
\begin{align*}
 U_C(\pi_m^\pm)
 &=(p^m)^{\text{\tiny{T}}}\phi p+(p^m)^{\text{\tiny{T}}}\psi p\pm\varepsilon\langle\phi,Z\rangle\\
 &=(p^m)^{\text{\tiny{T}}}(\phi+\psi)p\pm\varepsilon h\ \rightarrow\ p^{\text{\tiny{T}}}(\phi+\psi)p\pm\varepsilon h.
\end{align*}
Both $\phi$ and $\psi$ are skew-symmetric, so $p^{\text{\tiny{T}}}(\phi+\psi)p=0$. For all sufficiently large $m$, therefore, $U_C(\pi_m^+)>0$ and $U_C(\pi_m^-)<0$, meaning $\pi_m^+\in\widehat\Pi$ and $\pi_m^-\in\check\Pi$.

Completeness implies $(\pi_m^-)^{\text{\tiny{T}}}\in\Pi$. As both transposes lie in $\Gamma(p,p^m)$, correlation insensitivity at $p$ and $(\pi_m^-)^{\text{\tiny{T}}}\in\Pi$ force $(\pi_m^+)^{\text{\tiny{T}}}\in\Pi$. This contradicts $\pi_m^+\in\widehat\Pi$. Thus $C$ cannot be proper, so $C=\Delta$, and hence $\Delta^S=\Delta$.
\end{proof}

The representations obtained so far apply whenever the first marginal has full support. The following lemma, proved in Appendix \ref{appendix:boundary-extension}, extends them to boundary marginals.

\begin{lem}[Boundary Extension of CWU]\label{lem:cwu-boundary}
 Suppose a correlation preference satisfies continuity, completeness, and betweenness. Let $U(\pi):=\langle\phi,\pi\rangle+\pi_1^{\text{\tiny{T}}}\psi\pi_2$ for skew-symmetric matrices $\phi$ and $\psi$ . If $U$ represents weak preference and indifference whenever $\pi_1\in\Delta$, then it does so on all of $\Delta(X\times X)$.
\end{lem}

\begin{proof}[Proof of Theorem \ref{thm:cwu}]
Suppose continuity, completeness, and betweenness hold and the preference is correlation sensitive at some $p\in\Delta$. Theorem \ref{thm:dichotomy} rules out the insensitive branch and yields $\Delta^S=\Delta$, together with a nondegenerate CWU representation whenever the first marginal is interior. Lemma \ref{lem:cwu-boundary} extends that representation to the entire joint-distribution simplex.

Conversely, suppose $U(\pi)=\langle\phi,\pi\rangle+\pi_1^{\text{\tiny{T}}}\psi\pi_2$ is a nondegenerate CWU representation, normalized so that $\phi\notin\Sspace$. Skew symmetry gives $U(\pi^{\text{\tiny{T}}})=-U(\pi)$ and therefore completeness; continuity and fiberwise betweenness follow from continuity and linearity of $U$ on each fixed-first-marginal fiber. Since $\Sspace=\Tspace^\perp$, choose $Z\in\Tspace$ with $\langle\phi,Z\rangle\ne0$. For any $p\in\Delta$ and sufficiently small $\varepsilon>0$, both $pp^{\text{\tiny{T}}}\pm\varepsilon Z$ belong to $\Gamma(p,p)$ and have utilities $\pm\varepsilon\langle\phi,Z\rangle$. Hence the preference is correlation sensitive at every $p\in\Delta$.
\end{proof}

\subsection{The CI Branch and the Uniform Representation}\label{subsec:ssb-wu}

\begin{figure}[h]
\begin{center}

\begin{subfigure}[t]{0.38\textwidth}\centering
\begin{adjustbox}{max width=\linewidth}
\begin{tikzpicture}[scale=1.0, line cap=round]
\coordinate (Lbot) at (0,-2);
  \coordinate (Ltop) at (0,2);
  \coordinate (R)    at (5,0);    

  \draw[thick, dashed] (Lbot) -- (Ltop);

  \draw[red] (Ltop) -- node[midway, above right, sloped] {\scriptsize indifferent to $\bar p$} (R);
  \coordinate (Pa) at (0,0.8);
  \draw[red] (Pa)     -- (R);       
  \coordinate (Pb) at (0,-0.8);
  \draw[blue] (Pb)     -- (R);       
  \draw[red] (Lbot) -- node[midway, below right, sloped] {\scriptsize indifferent to $\underline p$} (R);

  \coordinate (Qt) at (2,1.2);
  \coordinate (Qb) at (2,-1.2);
  \draw[thick, dashed] (Qt) -- (Qb);

  \coordinate (QA) at (intersection of Qt--Qb and Pa--R);
  \coordinate (QB) at (intersection of Qt--Qb and Pb--R);

  \fill (QA) circle (1.3pt) node[above left, xshift=2pt] {$q_\alpha$};
  \fill (QB) circle (1.3pt) node[below left] {$q_\beta$};

  \node[left]  at (Ltop) {$\bar p$};
  \node[left]  at (Pa) {$p_\alpha$};
  \node[left]  at (Pb) {$p_\beta$};
  \node[left]  at (Lbot) {$\underline p$};
  \node[above, xshift=-6pt] at (Qt) {$\bar q$};
  \node[below, xshift=-6pt] at (Qb) {$\underline q$};
\end{tikzpicture}
\end{adjustbox}
\caption{Projective independence}\label{fig: PI}
\end{subfigure}\hfill%
\begin{subfigure}[t]{0.30\textwidth}\centering
\begin{adjustbox}{max width=\linewidth}
\begin{tikzpicture}[scale=1.0, line cap=round]
  \coordinate (P) at (-1.5,0);
  \coordinate (Q) at (0,2);
  \coordinate (R) at (1.5,0);
  \coordinate (B) at (-0.5,-1);

  \draw[thick] (P)--(Q)--(R)--cycle;
  \draw[red] (P)--(B);
  \draw[red] (Q)--(B);
   \draw[red] (R)--(B);

  \node[left]  at (P) {$p$};
  \node[left] at (Q) {$q$};
  \node[right] at (R) {$r$};
 \fill (B) circle (1.4pt) node[below, xshift=10pt] {$s=\alpha p+\beta q+\gamma r$};
\end{tikzpicture}
\end{adjustbox}
\caption{Acyclic preference}\label{fig: PI NT}
\end{subfigure}\hfill%
\begin{subfigure}[t]{0.28\textwidth}\centering
\begin{adjustbox}{max width=\linewidth}
\begin{tikzpicture}[scale=1.0, line cap=round]
  \coordinate (Q) at (0,1.5);
  \coordinate (P) at (-1.5,-1);
  \coordinate (R) at (1.5,-1);
  \coordinate (S) at (-0.2,0);

  \draw[thick] (P)--(Q)--(R)--cycle;

  \draw[red] (P)--(S);
  \draw[red] (Q)--(S);
  \draw[red] (R)--(S);

  \coordinate (QSext) at ($(Q)!2!(S)$);
  \coordinate (RSext) at ($(R)!2!(S)$);
  \coordinate (PSext) at ($(P)!2!(S)$);

  \coordinate (A) at (intersection of Q--QSext and P--R);
  \coordinate (B) at (intersection of R--RSext and P--Q);
  \coordinate (C) at (intersection of P--PSext and Q--R);

  \draw[red] (A)--(S);
  \draw[red] (B)--(S);
  \draw[red] (C)--(S);
  \fill (S) circle (1.4pt) node[below, xshift=2pt] {$s$};
  \node[left]  at (P) {$p$};
  \node[left] at (Q) {$q$};
  \node[right] at (R) {$r$};
\end{tikzpicture}
\end{adjustbox}
\caption{Cyclic preference}\label{fig: PI T}
\end{subfigure}
\caption{Implications of projective independence.}\label{fig: PIand projectivity}
    \end{center}
\begin{minipage}{\textwidth}
   \vspace{1ex} \footnotesize{Notes: Each red/blue line corresponds to an indifference line in a 2-dimensional probability simplex. The lines are extended affinely beyond the simplex; their common intersection need not be a lottery. In (a), $p_\alpha$ denotes $\alpha \bar p+(1-\alpha) \underline p$ for $\alpha\in(0,1)$, and likewise for $p_\beta,q_\alpha,q_\beta$. In (b), the preference is transitive for $\{p,q,r\}$ with $p\succ q$, $q\succ r$ and $p\succ r$.  In (c), the preference has a cycle $p\succ q$, $q\succ r$ and $r\succ p$.}
\end{minipage}
\end{figure}
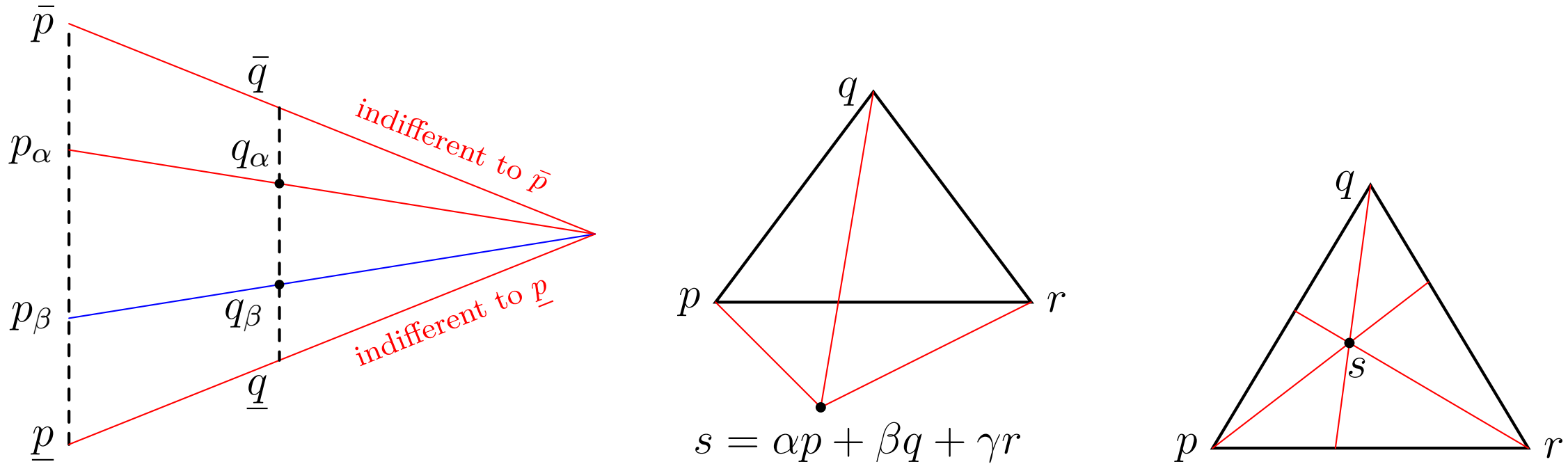

Theorem \ref{thm:dichotomy} leaves the globally insensitive branch as a correlation-insensitive CBU, or equivalently a conditional linear utility family. The following cross-fiber condition supplies its SSB normalization; it is not used in Theorems \ref{thm:dichotomy} or \ref{thm:cwu}. Projective independence is the restriction of the condition in \citet{chew1994projective} to correlation-insensitive comparisons. If two endpoint comparisons are indifferent and one interior mixture is also indifferent, the entire segment must be indifferent. Figure \ref{fig: PIand projectivity} illustrates both the axiom and the concurrence geometry used in the proofs.

\begin{axm}[Projective Independence (PI)]\label{axm:pi}
If $\bar p\sim^*\bar q$, $\underline p\sim^*\underline q$, and there is an $\alpha\in(0,1)$ such that $\alpha\bar p+(1-\alpha)\underline p\sim^*\alpha\bar q+(1-\alpha)\underline q$, then $\beta\bar p+(1-\beta)\underline p\sim^*\beta\bar q+(1-\beta)\underline q$ for every $\beta\in(0,1)$.
\end{axm}

\begin{defi}[Skew-Symmetric Bilinear Utility]\label{def:ssb}
The preference admits an {SSB representation} if there is a skew-symmetric matrix $\psi$ such that, for every $\pi\in\Delta(X\times X)$, $\pi\in\Pi$ if and only if $\pi_1^{\text{\tiny{T}}}\psi\pi_2\geq0$, with equality if and only if $\pi\in\widetilde\Pi$.
\end{defi}

SSB is CI by construction. The result below confirms it as the $\phi=0$ specialization of CWU. Appendix \ref{appendix: SSB} gives a self-contained proof parallel to the sensitivity-graph construction. Observe that PI provides the cycle-product consistency argument here, whereas the counterpart argument in the sensitive branch rests on correlation sensitivity (and completeness).
\begin{prop}[SSB Normalization of the Correlation-Insensitive Branch]\label{lem:ssb-branch}
For a preference satisfying continuity, completeness, and betweenness, if it is CI on $\Delta$ and satisfies projective independence, then the preference admits an SSB representation.
\end{prop}

\begin{defi}[Weighted Utility]\label{def:wu}
The preference admits a {weighted utility} representation if there are $a\in\R^n$ and $w\in\R_{++}^n$ such that, for every $\pi\in\Delta(X\times X)$, $\pi\in\Pi$ if and only if $\frac{a^{\text{\tiny{T}}}\pi_1}{w^{\text{\tiny{T}}}\pi_1}\geq\frac{a^{\text{\tiny{T}}}\pi_2}{w^{\text{\tiny{T}}}\pi_2}$, with equality if and only if $\pi\in\widetilde\Pi$. Equivalently, its has an SSB representation with matrix $\psi=aw^{\text{\tiny{T}}}-wa^{\text{\tiny{T}}}$.
\end{defi}

The following is the familiar connection between SSB and WU \citep{chew1983generalization,fishburn1983transitive}. Say the preference is \emph{CI-transitive} if $(p\succeq^*q$ and $q\succeq^*r)\implies p\succeq^*r$. 
\begin{prop}[SSB and WU]\label{prop:ssb-wu}
An SSB preference is CI-transitive if and only if it admits a WU representation.
\end{prop}
We record in Appendix \ref{appendix:rank} a concise alternative argument. CI-transitivity first forces $\operatorname{rank}(\psi)\leq2$. A rank-two skew-symmetric matrix has the form $aw^{\text{\tiny{T}}}-wa^{\text{\tiny{T}}}$; ordering the outcomes then permits a normalization with $w$ strictly positive. The preceding results now give a uniform representation of both branches.

\begin{prop}[Uniform Correlation Weighted Utility]\label{prop:cwu-uniform}
A correlation preference satisfies continuity, completeness, betweenness, and projective independence if and only if it admits a CWU representation. In the correlation-sensitive branch the representation is nondegenerate; in the correlation-insensitive branch it can be normalized to SSB utility.
\end{prop}

\begin{proof}
Suppose first that the four axioms hold. Theorem \ref{thm:dichotomy} gives two cases. If $\Delta^S=\Delta$, Theorem \ref{thm:cwu} supplies a nondegenerate global CWU representation. If $\Delta^I=\Delta$, Proposition \ref{lem:ssb-branch} supplies an SSB representation, which is CWU with $\phi=0$.

Conversely, let $U(\pi)=\langle\phi,\pi\rangle+\pi_1^{\text{\tiny{T}}}\psi\pi_2$ be a CWU representation. Skew symmetry gives $U(\pi^{\text{\tiny{T}}})=-U(\pi)$, implying completeness. Continuity and betweenness are immediate. Projective independence follows from bilinearity of SSB and linearity of CEU kernels. 
\end{proof}

\subsection{CEU and EU}\label{subsec: CEU details}

We now zoom in on the special case of CEU. The correlation analogue of Samuelson's four-lottery condition \citep{samuelson1952probability,fishburn1975separation} strengthens betweenness by allowing the first marginal to change.

\begin{axm}[Strong Independence]\label{axm:strong-independence}
For $\pi,\pi'\in\Delta(X\times X)$ and $\alpha\in(0,1)$:
\begin{enumerate}[noitemsep,topsep=2pt]
 \item $\pi,\pi'\in\widetilde\Pi$ implies $\alpha\pi+(1-\alpha)\pi'\in\widetilde\Pi$;
 \item $\pi\in\widehat\Pi$ and $\pi'\in\Pi$ implies $\alpha\pi+(1-\alpha)\pi'\in\widehat\Pi$;
 \item $\pi\in\check\Pi\cup\widetilde\Pi$ and $\pi'\in\check\Pi$ implies $\alpha\pi+(1-\alpha)\pi'\in\check\Pi$.
\end{enumerate}
\end{axm}

In the $\succ^\pi$ notation, the second clause says that $p\succeq^\pi q$ and $p'\succ^{\pi'}q'$ imply $p_\alpha\succ^{\pi_\alpha}q_\alpha$, where $\pi_\alpha:=\alpha\pi+(1-\alpha)\pi'$, $p_\alpha:=\alpha p+(1-\alpha)p'$, and $q_\alpha:=\alpha q+(1-\alpha)q'$. The indifference and rejection clauses have the analogous form. Unlike betweenness, strong independence imposes these restrictions even when $p\neq p'$. It therefore makes $\hat{\Pi}$ and $\check{\Pi}$ convex in the entire joint-distribution simplex rather than only fiber by fiber. The following lemma reduces CWU to CEU under strong independence via a perturbation argument.

\begin{lem}[Strong Independence Collapses CWU to CEU]\label{lem:cwu-ceu}
Suppose a preference admits a CWU representation
$U(\pi)=\langle\phi,\pi\rangle+\pi_1^{\text{\tiny{T}}}\psi\pi_2$. If the first clause of strong independence holds, then $\psi\in\Sspace$.
\end{lem}

Say the preference is \emph{reflexive} if every symmetric $\pi$ is indifferent. The following direct characterization is proved in Appendix \ref{appendix:ceu}.
\begin{prop}[Axiomatization of CEU]\label{prop:ceu}
A continuous correlation preference satisfies reflexivity and strong independence if and only if it admits a CEU representation.
\end{prop}

\begin{cor}[Axiomatization of CEU under Completeness]\label{cor:ceu-cs}
A correlation preference satisfies continuity, completeness, and strong independence if and only if it admits a CEU representation.
\end{cor}
Completeness implies reflexivity by Lemma \ref{lem:transpose}, so the corollary follows directly from the proposition. Lemma \ref{lem:cwu-ceu} separately locates CEU within CWU: strong independence removes the non-CEU marginal correction without requiring the projective normalization of the correlation-insensitive branch. 

The two axiomatic routes shown in Figure \ref{fig:routes-eu} can now be read formally. Along the right route, strong independence reduces CWU to CEU, and SML-CI at one full-support marginal makes the CEU kernel additively separable, yielding EU. Along the left, correlation insensitivity reduces CWU to SSB, and ordinary independence yields EU.

\section{Experimental Implications and a Joint Test}\label{section:experiments}

\subsection{Existing Experimental Designs}\label{subsec:existing-experiments}

Two experimental literatures study closely related manifestations of correlation sensitivity. The first uses SML comparisons, in which the alternatives have identical marginal distributions. Such comparisons appear in \citet{leland1998similarity} and are implemented directly by \citet{loewenfeld2025salience}. The latter identify a correlation-sensitive minority of about 37\%, with aggregate choices in the likelihood-sensitive rather than the canonical RT/ST contrast-sensitive direction. A similar pattern is found in \citet{bohren2024cognitive}, where they use same-distribution asset choices and differentiate how simultaneous and sequential exposure change subjects' mental representations and choices. 
The probability dominance theory of \citet{diecidue2020probability} provides an alternative explanation for the SML choice patterns. The probability-dominance criterion chooses the alternative that is more likely to yield the higher statewise payoff; in our language, it is CEU with kernel \(\phi^{\mathrm{PD}}(x,y)=\mathbf 1\{x>y\}-\mathbf 1\{x<y\}\). Their motivating equal-marginal comparison therefore belongs to the SML class: probability dominance favors the alternative that wins more often by smaller amounts rather than the alternative that wins less often by a larger amount. For any \(x_H>x_M>x_L\), \(\Delta_{\phi^{\mathrm{PD}}}=1-1-1=-1\), so probability dominance is DSOD and selects the DSOD alternative in our oriented SML task. 

The second literature varies the payoff correlation in the \citet{allais1953comportement} common-consequence problems while holding each alternative's marginal lottery fixed. One can consider the classical presentation as an independent coupling of the two marginals and Savage's statewise presentation as a highly positively correlated coupling of the same marginals. Marginal-only models predict no change between these presentations. A statewise comparison model could generally predict a change, because the pairs of obtained and forgone outcomes change even though the two lotteries do not. Table \ref{tab:allais-correlation-evidence} summarizes the standard-direction rates of Allais behavior reported in several implementations. 
The finding is qualitative: The rate of Allais behavior tends to be high in the standard setting of mutually independent lotteries and remains positive under strong positive dependence. \citet{frydman2018role}, for example, find a decline from 49\% under independence to 36\% under an intermediate coupling and 15\% under maximal positive dependence. \citet{bruhin2022risk} report 28\% and 17\% under their independent and dependent treatments.

\vspace{-2ex}
\begin{table}[h]
\begin{center}

\caption{Reported standard-direction Allais behavior across payoff couplings}
\label{tab:allais-correlation-evidence}
\small
\begin{tabularx}{\linewidth}{@{}>{\raggedright\arraybackslash}Xccc@{}}
\toprule
Study & Independent & Intermediate & Maximal/dependent\\
\midrule
\citet{frydman2018role} & 49\% & 36\% & 15\% \\
\citet{bruhin2022risk} & 28\% & --- & 17\% \\
\citet{loewenfeld2024event} & 62\% & --- & 18\% \\
\citet{starmer1992testing} & --- & --- & 33--36\% \\
\citet{bordalo2012salience} & --- & --- & 20\% \\
\citet{ostermair2022experimental} & --- & --- & 14--36\% \\
\citet{humphrey2023accepts} & --- & --- & 40\% \\
\bottomrule
\end{tabularx}
    
\end{center}
\vspace{1ex}\footnotesize \emph{Note:} Entries reproduce the studies' reported rates for their respective Allais-pattern definitions. Designs, stakes, samples, and the construction of dependence differ, so rows should not be compared as a pooled treatment effect.
\vspace{-2ex}
\end{table}
CWU separates the two empirical forces behind this pattern. Varying only the correlation changes $\langle\phi,\pi\rangle$ while leaving $\psi(p,q)$ fixed. The positive rate remaining under maximal correlation can therefore be carried by the marginal component, whereas the change across couplings isolates the corresponding correlation contrast. Pure CEU permits only the correlation effect; a correlation-insensitive SSB/WU model permits only the marginal residual.\footnote{Also note the event-splitting qualification: if a correlation treatment also changes the number or visual partition of repeated outcomes, the design no longer changes only $\pi$ \citep{loewenfeld2024event}.}

Existing SML and correlation Allais studies nevertheless remain separate in the sense relevant to Theorem \ref{thm:dichotomy}. They use different tasks, supports, or type classifications and do not ask whether one's local SML response and cross-coupling Allais pattern satisfy the common-kernel restriction. The following design administers both tasks to the same subjects on a common outcome support. 

\subsection{A Joint SML--HILO Test}\label{subsec:joint-test}

The predictions in Section \ref{subsec:xsod} suggest combining the same-marginal design of \citet{loewenfeld2025salience} with the four HILO situations. Eliciting both blocks from the same subjects on a common support $x_L<x_M<x_H$ tests whether one correlation kernel accounts for their local SML responses and their Allais responses across couplings, and separating these effects from the marginal CWU term in the meantime.

The SML block uses the three equally likely payoff pairs $(x_L,x_M)$, $(x_M,x_H)$, and $(x_H,x_L)$. Both alternatives have the same uniform marginal, so the marginal term vanishes and the comparison has value $\Delta_\phi/3$, where $\Delta_\phi=\phi(x_H,x_L)-\phi(x_H,x_M)-\phi(x_M,x_L)$.

The HILO block starts from $A_O=[x_M]$ and $B_O=\beta[x_H]+(1-\beta)[x_L]$, with $\beta\in(0,1)$. Mixing each alternative with the same consequence $x_M$, $x_L$, or $x_H$, using common-consequence probability $1-\alpha\in(0,1)$, gives the $I$-, $L$-, and $H$-situations in Table \ref{tab:hilo-marginals}. Within each situation, use the coupling path $\pi_z^\lambda=(1-\lambda)\pi_z^{\mathrm{ind}}+\lambda\pi_z^+$ at $\lambda\in\{0,\tfrac12,1\}$, where the endpoints are independence and comonotonicity. The $O$- and $I$-situations are coupling invariant because $A$ is degenerate. In the $L$- and $H$-situations, equations \eqref{eq:xsod-allais-L}--\eqref{eq:xsod-allais-H} link treatment effects to the SML index. As $\lambda$ falls, local ISOD predicts movement toward $B_L$ and $A_H$, local DSOD predicts the reverse movement, and local CSOD predicts no movement. 

The maximal-correlation arm identifies residual marginal effects. Writing $C_z^\lambda:=\langle\phi,\pi_z^\lambda\rangle$, one has $C_L^1=C_I^1=C_H^1$: the non-tie statewise comparisons are identical, while the remaining probability lies on diagonal payoff pairs. Pure CEU therefore cannot produce an $I$--$L$ or $I$--$H$ reversal in this arm. Under CWU, such a reversal can arise from differences among $\psi(A_I,B_I)$, $\psi(A_L,B_L)$, and $\psi(A_H,B_H)$. Coupling changes thus isolate the correlation effect, while reversals remaining at maximal correlation requires a marginal contribution.

Let $S_i$ be a signed SML measure, positive for repeated ISOD choices. A parsimonious latent-index specification for $z\in\{L,H\}$ is $$A_{iz\lambda}^*=\mu_{iz}+\tau_z(1-\lambda)+\kappa_z(1-\lambda)S_i+\varepsilon_{iz\lambda},$$where $A_{iz\lambda}=\mathbf 1\{A_{iz\lambda}^*\geq0\}$ indicates choice of $A$. The matched reduced-form prediction is $\kappa_L<0<\kappa_H$, provided $S_i$ is a monotone proxy for $\Delta_{\phi_i}$ on a common latent scale and the response mechanism is stable across treatments. 

\subsection{Distinguishing CWU, CEU, SSB, and EU Specifications}\label{subsec:model-types}

The design can also estimate behavioral types rather than only average treatment effects. \citet{bruhin2022risk} illustrate the value of such out-of-sample fitting and separate EU, probability-weighting, and salience types. Here. within CWU, CEU requires $\psi\in\Sspace$ being separable; sensitive CEU additionally has $\phi\neq0$; SSB has $\phi=0$; and EU has $\phi=0$ and $\psi\in\Sspace$. For instance, On a three-point support these conditions are respectively $\Delta_\psi=0$, $\Delta_\phi\neq 0$, $\Delta_\phi=0$, and both indices zero. A disjoint exhaustive partition within CWU is therefore EU, SSB outside EU, sensitive CEU, and sensitive CWU outside CEU. For brevity, we abuse the terminology and call the four groups EU, SSB, CEU and CWU groups, respectively.

The above tasks already provide distinct sources of identification. Coupling changes at fixed marginals isolate correlation contrasts, and on the three-outcome support the SML cycle identifies the sign of $\Delta_\phi$. HILO patterns at maximal correlation identify particular marginal contrasts, while conventional independent-lottery mixture tasks help distinguish SSB from EU. Thus, the SML test distinguishes the first two groups from the last two, while the HILO patterns further separates the expected utility groups (EU-CEU) from the non-expected utility groups (SSB-CWU), constituting a two-by-two categorization.

A stochastic specification can further assign type $k$ the choice probability $\Pr(A\mid\pi,k)=\Lambda(\sigma_k U_k(\pi))$ and estimate the type weights jointly with the parsimonious kernel parameters. At a fixed $\beta$, however, the displayed SML--HILO block does not identify every normalized CWU coefficient. With $\Delta_\psi:=\psi(x_H,x_L)-\psi(x_H,x_M)-\psi(x_M,x_L)$, one can write all the latent utilities as functions of $U_O$, $\Delta_\psi$ and $\Delta_\phi$. Since the CWU kernels contain $4$ parameters on the $LMH$ outcome space after all normalization (1 for $\phi$, 3 for $\psi$), full kernel estimation therefore requires a second $\beta'\neq\beta$ or another marginal comparison. 

\section{Discussion}\label{section:discussion}

\subsection{Counterfactual Utility and Non-Binary CEU}
\label{subsec:kis-context-projection}

The local-to-global implication extends to choice among multiple alternatives. We discuss a generalization of CEU that compares multiple correlated lotteries building on the context-dependent counterfactual utility of \citet{koch2026counterfactual} (hereafter KIS). We show that indifference among all decisions under every coupling of \emph{one} common full-support marginal makes choice invariant to correlation across all marginal profiles. The conclusion holds when comparisons are allowed to depend on other alternatives’ marginal distributions. The extension therefore distinguishes sensitivity to correlation from dependence on the marginal context.

Suppressing covariates and using our finite outcome space $X$, let $D=\{0,\ldots,K-1\}$ index the decisions and let $\mathbf Y=(Y(k))_{k\in D}$ be their potential outcomes. A joint distribution $P\in\Delta(X^D)$ specifies how these outcomes are jointly realized. The utility $\tilde u(d;\mathbf y)$ may depend on both the realized outcome $y_d$ and all forgone outcomes. KIS assign each decision a value and choose from a nonempty submenu $A\subseteq D$ according to
\begin{equation}\label{eq:kis-context-choice}
 V_d(P):=\mathbb E_P[\tilde u(d;\mathbf Y)],
 \qquad
 C_P(A):=\mathop{\rm arg\,max}_{d\in A}V_d(P).
\end{equation}

To study correlation preference, we hold $D$ and $\tilde u$ fixed while varying $P$. This construction gives a model of choice from menus whose binary specialization, under the relabeling condition below, is CEU. For $D=\{0,1\}$, let $\phi(x,y):=\tilde u(0;x,y)-\tilde u(1;x,y)$. Under the relabeling condition $\tilde u(0;x,y)=\tilde u(1;y,x)$, the kernel $\phi$ is skew-symmetric and $0\succeq^P1\iff\mathbb E_P[\phi(Y(0),Y(1))]\geq0$, exactly as under CEU. 

For $|D|\ge 3$, comparisons may depend on the full context. Let $\Gamma_D(\mathbf p)$ be the set of joint distributions with marginals $\mathbf p=(p_k)_{k\in D}$. The natural extension of SML-CI at $p\in\Delta$ requires $C_P(D)=D$ for every $P\in\Gamma_D(p,\ldots,p)$: all decisions are indifferent under every coupling of the common marginal. Global CI requires $C_P(A)=C_Q(A)$ for every marginal profile $\mathbf p$, every $P,Q\in\Gamma_D(\mathbf p)$, and every nonempty $A\subseteq D$. The following local-to-global result uses the multi-marginal version of the OT argument; its proof is in the appendix.

\begin{prop}[SML-CI under Context-Dependent Projection]
\label{prop:kis-sml-ci}
If the choice rule in \eqref{eq:kis-context-choice} is SML-CI at some full-support marginal $p^*\in\Delta$, there exist functions $h:X^D\to\mathbb R$ and $a_{dk}:X\to\mathbb R$ such that $\tilde u(d;\mathbf y)=h(\mathbf y)+\sum_{k\in D}a_{dk}(y_k)$ for every $d\in D$ and $\mathbf y\in X^D$. Consequently, choices are globally CI with $C_P(A)=\mathop{\rm arg\,max}_{d\in A}\sum_{k\in D,x\in X}a_{dk}(x)p_k(x)$.
\end{prop}
To obtain the full dichotomy, suppose additionally that all decisions are indifferent whenever their potential outcomes coincide almost surely: $ \tilde u(i;x,\ldots,x)=\tilde u(j;x,\ldots,x)$ for every $x\in X$ and $i,j\in D$. Under this restriction, the coupling $\sum_{x\in X}p(x)\delta_{(x,\ldots,x)}$ makes all decisions indifferent. Global CI then extends this indifference to every coupling in $\Gamma_D(p,\ldots,p)$. Together with Proposition~\ref{prop:kis-sml-ci}, this establishes
\[
 \text{SML-CI at some }p\in\Delta
 \quad\Longleftrightarrow\quad
 \text{global CI}
 \quad\Longleftrightarrow\quad
 \text{SML-CI at every }p\in\Delta.
\]
Without the diagonal indifference, the proposition's forward implication remains valid, but global CI need not imply same-marginal indifference due to labeling. For $|D|\ge 3$, the marginal expression for $V_i(P)-V_j(P)$ may involve $p_k$ for $k\notin\{i,j\}$. Correlation insensitivity can therefore coexist with dependence on the other options' marginal lotteries. 

\subsection{A Copula-Dual Utility}\label{subsec:copula-dual}

To explore the scope of the dichotomy, we consider a correlation-sensitive analogue of Yaari’s dual theory \citep{yaari1987dual}. Choice can depend on the alternatives’ copula, which records how their outcome ranks are paired. On an atomless domain, the construction below permits correlation effects to vanish at one full-support marginal and remain active at another, as betweenness is not imposed. 

For this subsection, let $X=[0,1]$ and restrict attention to atomless full-support distributions. For $p$, define the survival function $G_p(x):=\Pr_p\{t>x\}$ and its inverse, the decreasing quantile function $Q_p(u):=\inf\{x:G_p(x)\leq u\}$. Small $u$ denotes a high outcome, and $d\mu_p:=-dQ_p$ is a probability measure describing outcome increments across ranks. Yaari's dual utility is $D_f(p):=\int_0^1 f(u)d\mu_p(u)$, where $f$ is continuous and increasing with $f(0)=0$ and $f(1)=1$.

For $(t_1,t_2)\sim\pi\in\Gamma(p,q)$, let $c_\pi$ be the copula of $(G_p(t_1),G_q(t_2))$. Drawing a rank pair from this copula and applying $(Q_p,Q_q)$ recovers the joint distribution $\pi$; see \citet{nelsen2006introduction}. For bounded Borel-measurable rank-pair kernels $\phi^c$ indexed by the copula, consider
\begin{equation}
 U_\phi(\pi):
 =\int_{[0,1]^2}\phi^{c_\pi}(u,v)d\mu_p(u)d\mu_q(v),
 \label{eq:cdu-general}
\end{equation}
with $\pi\in\Pi\iff U(\pi)\ge 0$ as before. Correlation changes the kernel, while the marginals determine the averaging weights. Imposing reciprocity, $\phi^{c^{\text{\tiny T}}}(v,u)=-\phi^c(u,v)$, gives $U_\phi(\pi^{\text{\tiny T}})=-U_\phi(\pi)$ and completeness. 

The specialization $\phi^c(u,v)=f(u)-f(v)+\varphi(c)$ gives $U_{f,\varphi}(\pi)=D_f(p)-D_f(q)+\varphi(c_\pi)$ with $\varphi(c^{\text{\tiny T}})=-\varphi(c)$. The preference agrees with Yaari on comonotonic pairs and satisfies \textit{comonotonic independence}: mixing pairwise comonotonic prospects state by state with the same common prospect preserves ranking due to linearity of $D_f$ in the quantiles.\footnote{Intuitively, it adds a copula premium to the Yaari comparison: the comonotonic copula $c_{min}(u,v)=\min\{u,v\}$ is symmetric, so its premium vanishes. } The general form \eqref{eq:cdu-general} need not exhibit the bang-bang property: a copula effect can cancel under one marginal's averaging weights without cancelling under all. The following proposition provides one such instance and is proved in the appendix.
\begin{prop}\label{prop: CDU CI}
    Let $h(u)=2u-1$ and $\psi\not\equiv 0$ be continuous and satisfy reciprocity. For a copula dual utility with $\phi^c(u,v)=f(u)-f(v)+h(u)h(v)\varphi(c)$, it is SML-CI at the uniform marginal but not CI in the entire domain.
\end{prop}

\subsection{Optimal Transport and Graph-Theoretic Integration}\label{subsec:discussion-ot}

The proof combines two forms of linear duality. Fiberwise separation turns betweenness into local hyperplanes, while Kantorovich duality characterizes the matrices whose expectation is invariant across a transportation polytope. In the latter step, the row and column potentials are shadow prices on the marginal constraints; additive separability means that all apparent joint-state variation can be priced through the marginals alone. This recasts the EU kernel as the correlation-insensitive part of a comparative evaluation in the form $\phi(x,y)=u(x)-u(y)$. Related OT ideas have proved useful in multivariate risk and other economic-design problems \citep{galichon2012dual,lin2022credible,corrao2023randomization,kolotilin2025persuasion}.

The graph step solves a different problem: initially, local separating kernels are cardinally unrelated across fibers. Completeness supplies an edge scale for every correlation-sensitive pair, and the product-one restriction around each cycle makes these scales integrable. A potential then normalizes the local kernels, much as cycle conditions and dual prices do in network-flow and incentive-compatibility arguments \citep{vohra2011mechanism}. The method is useful here because it does not commit to a parametric regret or salience function. RT and ST enter through different interpretations and restrictions on the common kernel $\phi$, while the sensitivity graph provides the global consistency needed before either interpretation can be tested across lottery pairs.

\section{Conclusion}\label{section:conclusion}

The present paper studies correlation preferences exclusively under risk, focusing on the restrictions imposed by continuity, completeness, and betweenness. We show that correlation insensitivity is either void or global under the axioms. Our companion paper \citep{chenChewZhang2026CPS}, inheriting content not included here from an earlier version of this draft, starts from the other perspective in a broader correlation preference framework. There, we develop a Savagean framework in which correlation probabilistic sophistication makes joint outcome distributions sufficient for binary choice, extending probabilistic sophistication \citep{machina1992more} beyond the correlation-insensitive preference domain. Connection to CWU, CEU, and SSB are discussed.

\begingroup
\singlespacing
\small
\setlength{\bibsep}{0.6\baselineskip}
\bibliographystyle{apalike}
\bibliography{ref}
\endgroup

\appendix

\section{Extension to a General Outcome Space}
\label{sec:finite-support-extension}

We now allow an infinite outcome space while retaining finitely supported lotteries as in \citet{lanzani2022correlation}. Let $X$ be a metrizable outcome space. We retain only finitely supported probability measures and work with the family of finite-support simplexes. Proofs are collected in Appendix \ref{appendix:finite-support-proofs}
.

\subsection{Domain, topology, and behavioral conditions}

Let $\Pf(X)$ be the set of finitely supported Borel probability measures on $X$, and let $\Jf(X):=\Pf(X^2)$. Both sets carry the relative weak topology inherited from the corresponding spaces of Borel probability measures. 
Continuity below therefore applies to weakly convergent sequences whose limits lie in $\Jf(X)$. 

For $\pi\in\Jf(X)$, write $\pi_1$ and $\pi_2$ for its marginals and $\pi^\T$ for its transpose. For $p,q\in\Pf(X)$, set as before $\Gamma_f(p,q):=\{\pi\in\Jf(X):(\pi_1,\pi_2)=(p,q)\}$. A correlation preference is a set $\Pi\subseteq\Jf(X)$. Define $ \widehat\Pi$, $\widetilde\Pi$, and $\check\Pi:=\Jf(X)\setminus\Pi$ as in the finite setting.
Continuity, completeness, and betweenness are restricted to $\Jf(X)$ correspondingly. For a nonempty finite set $S\subset X$, let
\[
 \Delta(S):=\{p\in\Pf(X):\supp(p)\subseteq S\},\qquad
 \Delta^\circ(S):=\{p\in\Pf(X):\supp(p)=S\},
\]
and $\mathcal J(S):=\{\pi \in J_f(X): \supp(pi)\subseteq S \times S\}$. The restriction $\Pi_S:=\Pi\cap\mathcal J(S)$ is a correlation preference on an ordinary finite-dimensional simplex and inherits all three axioms. On each finite face $\mathcal J(S)$, the relative weak topology coincides with the usual Euclidean topology.

\subsection{Support-lattice dichotomy}


The relevant counterpart of correlation insensitivity at $p$ is support-relative. For $p\in\Delta^\circ(S)$, say that the preference is \emph{$S$-CI at $p$} if membership in $\Pi$ is constant on $\Gamma_f(p,q)$ for every $q\in\Delta(S)$. Equivalently, this is ordinary correlation insensitivity at $p$ within the finite face $\mathcal J(S)$. Define the families of sensitive and insensitive supports by
\begin{equation*}
\begin{aligned}
 \mathfrak S&:=\{S\subset X:0<|S|<\infty,\ 
       \text{some SML-sensitive }p\in\Delta^\circ(S)\},\\
 \mathfrak I&:=\{S\subset X:0<|S|<\infty\}\setminus\mathfrak S.
\end{aligned}
\end{equation*}

\begin{thmbis}{thm:dichotomy}[Support-Lattice Correlation Dichotomy]
\label{thm:support-lattice}
Under continuity, completeness, and betweenness, for every nonempty finite $S\subseteq X$, exactly one of the following holds:
\begin{enumerate}[label=\textup{(\roman*)},topsep=3pt]
 \item $S\in\mathfrak I$. Every $p\in\Delta^\circ(S)$ is SML-CI and the preference is $S$-CI at every such $p$.
 \item $S\in\mathfrak S$. Every $p\in\Delta^\circ(S)$ is SML-sensitive. 
\end{enumerate}
Moreover, for nonempty finite $S\subseteq T\subseteq X$, $S\in\mathfrak S$ implies $T\in\mathfrak S$, and $T\in\mathfrak I$ implies $S\in\mathfrak I$.
\end{thmbis}
Insensitivity is relative to $S$: it restricts comparisons with lotteries supported on $S$, but does not rule out sensitivity if the outcome support is enlarged. 

\subsection{Global functional CWU}

Let $\Mf(X)$ and $\Mf(X^2)$ denote the vector spaces of finitely supported finite signed measures. For $Z\in\Mf(X^2)$, let $Z_1,Z_2$ denote its signed marginals. A linear functional $L:\Mf(X^2)\to\R$ is \emph{skew} if $L(Z^\T)=-L(Z)$. A bilinear functional $B:\Mf(X)\times\Mf(X)\to\R$ is \emph{skew} if $B(r,s)=-B(s,r)$. The linear functional $L$ is \emph{degenerate} if $L(Z)=0$ whenever $Z_1=Z_2=0$; otherwise it is nondegenerate. Thus $L$ is degenerate exactly when it cannot distinguish two couplings with the same marginals. For a skew $L$, degeneracy is equivalent to the existence of a function $u:X\to\R$ such that, on probability measures, $ L(\pi)=\int u\,d\pi_1-\int u\,d\pi_2$. Such a term can be absorbed into the bilinear part.

\begin{defi}[Functional CWU]
\label{def:functional-cwu}
A {functional CWU representation} is a pair $(L,B)$ consisting of a skew linear $L$ and a skew bilinear $B$ such that
\[
 U(\pi):=L(\pi)+B(\pi_1,\pi_2),\quad
 \Pi=\{\pi\in\Jf(X):U(\pi)\geq0\},\quad
 \widetilde\Pi=\{\pi:U(\pi)=0\}.
\]
\end{defi}
The CWU is \emph{weakly closed} if its nonnegative set is closed relative to $\Jf(X)$, and it is \emph{nondegenerate} if $L$ is nondegenerate. Every such functional has an equivalent kernel notation. Define $\phi(x,y):=L(\delta_{(x,y)})$ and $\psi(x,y):=B(\delta_x,\delta_y)$. Then $\phi$ and $\psi$ are skew functions on $X^2$ and, because every measure under consideration has finite support,
\begin{equation}
 U(\pi)
 =\sum_{x,y}\phi(x,y)\pi(x,y)
  +\sum_{x,y}\psi(x,y)\pi_1(x)\pi_2(y).
 \label{eq:functional-cwu-kernel}
\end{equation}

\begin{thmbis}{thm:cwu}[Global Functional CWU Representation]
\label{thm:direct-limit-cwu}
A correlation preference on $\Jf(X)$ satisfies continuity, completeness, and betweenness, and is SML-sensitive at some $p^0\in\Pf(X)$ if and only if it admits a weakly closed, nondegenerate functional CWU representation on all of $\Jf(X)$.
\end{thmbis}


\section{Proofs}\label{appendix:ceu}\label{appendix:fiber}\label{appendix:local-sensitivity}\label{appendix:component-cwu}\label{appendix:boundary-extension}

\begin{lem}[Transposition under Completeness]\label{lem:transpose}
If the preference is complete, then $\check\Pi=(\widehat\Pi)^{\text{\tiny{T}}}$. Consequently, $\pi\in\widehat\Pi$ if and only if $\pi^{\text{\tiny{T}}}\in\check\Pi$, and $\pi\in\widetilde\Pi$ if and only if $\pi^{\text{\tiny{T}}}\in\widetilde\Pi$. Every symmetric joint distribution is indifferent, and both $\widehat\Pi$ and $\check\Pi$ are open under continuity.
\end{lem}

\begin{proof}
If $\pi\in\check\Pi$, completeness gives $\pi^{\text{\tiny{T}}}\in\Pi$. Since $(\pi^{\text{\tiny{T}}})^{\text{\tiny{T}}}=\pi\notin\Pi$, one has $\pi^{\text{\tiny{T}}}\in\widehat\Pi$, and hence $\pi\in(\widehat\Pi)^{\text{\tiny{T}}}$. The converse follows directly from the definition of $\widehat\Pi$. The remaining equivalences follow because transposition is an involution. A symmetric joint distribution cannot be rejected under completeness and is therefore indifferent. Finally, $\check\Pi$ is open when $\Pi$ is closed, and $\widehat\Pi=(\check\Pi)^{\text{\tiny{T}}}$ is open because transposition is a homeomorphism.
\end{proof}

\begin{proof}[Proof of Proposition \ref{prop:cbu}]
Fix $p\in\Delta X$. For any $A\subseteq\Delta(X\times X)$, write $A|_p:=A\cap\Delta_p$. Betweenness makes both $\Pi|_p$ and $\check\Pi|_p$ convex: two weakly accepted points are covered by its first two clauses, and two rejected points by its third clause. Continuity makes $\Pi|_p$ closed and $\check\Pi|_p$ relatively open. Completeness guarantees that the symmetric distribution $pp^{\text{\tiny{T}}}$ belongs to $\widetilde\Pi|_p$, and Lemma \ref{lem:transpose} makes $\widehat\Pi|_p$ relatively open.

Betweenness also makes $\widehat\Pi|_p$, $\check\Pi|_p$, and $\widetilde\Pi|_p$ convex. Suppose first that $\widehat\Pi|_p$ and $\check\Pi|_p$ are both nonempty. Weakly separate these disjoint convex sets by a nonconstant affine functional $L$ on $\Delta_p$, oriented so that $L\geq0$ on $\widehat\Pi|_p$ and $L\leq0$ on $\check\Pi|_p$. Each set is relatively open and hence contains a relatively open subset of the affine hull of $\Delta_p$; therefore $L$ is positive somewhere on $\widehat\Pi|_p$ and negative somewhere on $\check\Pi|_p$. If $L$ vanished at a point of $\widehat\Pi|_p$, mixing that point slightly toward a negative point of $\check\Pi|_p$ would remain in $\widehat\Pi|_p$ by openness but have negative value. Thus $L>0$ on $\widehat\Pi|_p$; analogously, $L<0$ on $\check\Pi|_p$. For $\pi\in\widetilde\Pi|_p$, mixtures of $\pi$ with points of $\widehat\Pi|_p$ are strict, while mixtures with points of $\check\Pi|_p$ are rejected. Taking limits toward $\pi$ gives $L(\pi)\geq0$ and $L(\pi)\leq0$, so $L(\pi)=0$.

If $\widehat\Pi|_p$ is empty and $\check\Pi|_p$ is nonempty, then $\Pi|_p=\widetilde\Pi|_p$. The third clause of betweenness makes $\widetilde\Pi|_p$ a face of the polytope $\Delta_p$: an interior mixture in $\widetilde\Pi|_p$ cannot have a rejected endpoint. Every face of a polytope is exposed, so there is an affine functional that is zero exactly on $\widetilde\Pi|_p$ and negative on $\check\Pi|_p$. If $\check\Pi|_p$ is empty, the second clause similarly makes $\widetilde\Pi|_p$ a face; choose an affine functional zero on $\widetilde\Pi|_p$ and positive on $\widehat\Pi|_p$, taking it to be zero when $\widetilde\Pi|_p=\Delta_p$. Representing the resulting affine functional by a matrix $\phi_p$ and repeating the construction on every fiber gives CBU with the exact positive, zero, and negative regions.

For the converse, a CBU representation is linear on each fixed-first-marginal fiber. The three sign implications in betweenness follow immediately from linearity.
\end{proof}

\begin{proof}[Proof of Lemma \ref{lem:local-sensitivity}]
Since $p\in\Delta^S$, some $\rho\in\Gamma(p,p)$ is nonindifferent. Transposing if necessary, take $\rho\in\widehat\Pi$. The symmetric full-support distribution $pp^{\text{\tiny{T}}}$ is indifferent. For any $\eta\in (0,1)$, betweenness gives $\bar\rho^+=(1-\eta)\rho+\eta pp^{\text{\tiny{T}}}\in\widehat\Pi$. It has full support, while $\bar\rho^-=(\bar\rho^+)^{\text{\tiny{T}}}$ belongs to $\check\Pi$ and also has full support. 

For $r,s$ close to $p$, define $\bar\rho^\pm_{r,s}:=\bar\rho^\pm+rs^{\text{\tiny{T}}}-pp^{\text{\tiny{T}}}$. These matrices have marginals $(r,s)$. Full support of $\bar\rho^\pm$ ensures nonnegativity for $(r,s)$ sufficiently close to $(p,p)$, and $\bar\rho^\pm_{r,s}\to\bar\rho^\pm$. Openness of $\widehat\Pi$ and $\check\Pi$ then gives $\bar\rho^+_{r,s}\in\widehat\Pi$ and $\bar\rho^-_{r,s}\in\check\Pi$ throughout a sufficiently small neighborhood $N_p$. Hence every pair in $N_p\times N_p$ is correlation sensitive, and in particular every $r\in N_p$ belongs to $\Delta^S$.

An open connected subset of the affine hull of a finite-dimensional simplex is path connected. Given a path in one component, its compact image is covered by the local sensitivity neighborhoods. A Lebesgue number for a finite subcover permits a fine finite subdivision of the path such that each two successive points lie in one common neighborhood and hence form a sensitivity edge. This gives a finite edge path between the endpoints.
\end{proof}

\begin{proof}[Proof of Proposition \ref{prop:component-cwu}]
Take the normalized kernels $H_p$ and their common quotient class $h$ from Lemma \ref{lem:pairwise-scale}. Fix $p_0\in C$ and set $\phi_C:=\frac12(H_{p_0}-H_{p_0}^{\text{\tiny{T}}})$. Then $\phi_C$ is skew-symmetric and $[\phi_C]=h$. Hence, for each $p\in C$, there exist $u_p,w_p\in\R^n$ such that $H_p=\phi_C+u_p\one^{\text{\tiny{T}}}-\one w_p^{\text{\tiny{T}}}$. For $\pi\in\Gamma(p,q)$,
\[
 \langle H_p,\pi\rangle=\langle\phi_C,\pi\rangle+u_p\cdot p-w_p\cdot q
 =\langle\phi_C,\pi\rangle+v_p\cdot q,
\]
where $v_p:=(u_p\cdot p)\one-w_p$. The vector $v_p$ does not depend on the particular separable decomposition.

For each sensitivity edge $(p,q)$, the generalized skew-symmetry identity in Lemma \ref{lem:pairwise-scale} gives $H_p+H_q^{\text{\tiny{T}}}\in\Sspace$, with zero expectation on $\Gamma(p,q)$. The preceding local representation and skew symmetry of $\phi_C$ therefore give $v_p\cdot q+v_q\cdot p=0$ on every sensitivity edge $(p,q)$.

Fix $r\in C$ and take a sensitivity neighborhood $N_r$ from Lemma \ref{lem:local-sensitivity}. Choose $n$ linearly independent lotteries $q^1,\ldots,q^n\in N_r$, and let $Q$ be the invertible matrix with these lotteries as columns. For $p\in N_r$, the edge identities give
\[
 Q^{\text{\tiny{T}}}v_p=-(v_{q^1}^{\text{\tiny{T}}}p, \dots, v_{q^n}^{\text{\tiny{T}}}p)^{\text{\tiny{T}}}.
\]
The right-hand side is linear in $p$, so there is a matrix $\psi_r$ such that $v_p=\psi_r^{\text{\tiny{T}}}p$ on $N_r$. Substitution into the edge identity yields $p^{\text{\tiny{T}}}(\psi_r+\psi_r^{\text{\tiny{T}}})q=0$ for all $p,q\in N_r$. Since a relatively open subset of the simplex linearly spans $\R^n$, $\psi_r$ is skew-symmetric.

The local matrix is unique: if $\psi_r'$ has the same property, then $(\psi_r-\psi_r')^{\text{\tiny{T}}}p=0$ on a spanning set. Matrices associated with overlapping sensitivity neighborhoods therefore coincide. A finite chain of overlapping neighborhoods along a path in $C$ patches them into one skew-symmetric matrix $\psi_C$ with $v_p=\psi_C^{\text{\tiny{T}}}p$ for every $p\in C$. Consequently, $\langle H_p,\pi\rangle=\langle\phi_C,\pi\rangle+p^{\text{\tiny{T}}}\psi_Cq$ for every $p\in C$, every $q\in\Delta X$, and every $\pi\in\Gamma(p,q)$. The scope in $q$ is unrestricted because $v_p=\psi_C^{\text{\tiny{T}}}p$ is a vector identity. Since $H_p$ is a positive rescaling of the  local kernel $\phi_p$, this proves the claimed  representation. Finally, $\phi_C\notin\Sspace$; otherwise the separable decomposition of $H_p$ would make every $\phi_p$ separable, contrary to $C\subseteq\Delta^S$.
\end{proof}

\begin{proof}[Proof of Lemma \ref{lem:cwu-boundary}]
The representation is first seen to be exact whenever either marginal is interior. If $\pi_2\in\Delta$, it applies to $\pi^{\text{\tiny{T}}}$. Skew symmetry gives $U(\pi^{\text{\tiny{T}}})=-U(\pi)$, while Lemma \ref{lem:transpose} reverses strict and rejected comparisons and preserves indifference. Hence the asserted sign classification also holds when $\pi_2\in\Delta$.

Now let $p:=\pi_1$ and $q:=\pi_2$, allowing both to lie on the boundary. Suppose first that $U(\pi)<0$ and, toward a contradiction, that $\pi\in\Pi$. For $t\in(0,1)$, set $\sigma_t:=(1-t)pp^{\text{\tiny{T}}}+t\pi$ and $r_t:=(1-t)p+tq$. Since $pp^{\text{\tiny{T}}}$ is indifferent, betweenness implies $\sigma_t\in\Pi$. Let $a_t:=\sigma_t^{\text{\tiny{T}}}$ and $b_t:=(1-t)\pi+tqq^{\text{\tiny{T}}}$. Separate affinity of $U$ on fixed-first- and fixed-second-marginal fibers gives $U(a_t)=-tU(\pi)>0$ and $U(b_t)=(1-t)U(\pi)<0$.

Fix $s\in\Delta$. For sufficiently small $\varepsilon>0$, define $a_t^\varepsilon:=(1-\varepsilon)a_t+\varepsilon r_ts^{\text{\tiny{T}}}$ and $b_t^\varepsilon:=(1-\varepsilon)b_t+\varepsilon r_ts^{\text{\tiny{T}}}$. These two distributions have the same first marginal $r_t$, their second marginals are interior, and their utilities retain the signs of $U(a_t)$ and $U(b_t)$. The first paragraph therefore gives $a_t^\varepsilon\in\widehat\Pi$ and $b_t^\varepsilon\in\check\Pi$. Since $a_t^\varepsilon\to a_t$, continuity gives $a_t\in\Pi$.

The distribution $a_t$ cannot be indifferent. Otherwise, the third clause of betweenness would make every interior mixture of $a_t$ and $b_t^\varepsilon$ rejected. Such a mixture has an interior second marginal, whereas mixtures sufficiently close to $a_t$ retain positive utility and hence are strict by the first paragraph, a contradiction. Thus $a_t$ is strict, so Lemma \ref{lem:transpose} makes $\sigma_t=a_t^{\text{\tiny{T}}}$ rejected, contradicting $\sigma_t\in\Pi$. We have proved that $U(\pi)<0$ implies $\pi\in\check\Pi$. Applying this implication to $\pi^{\text{\tiny{T}}}$ shows that $U(\pi)>0$ implies $\pi\in\widehat\Pi$.

It remains to consider $U(\pi)=0$. Suppose that $\pi$ is strict. Betweenness makes $\sigma_t:=(1-t)pp^{\text{\tiny{T}}}+t\pi$ strict for every $t\in(0,1)$. With $r_t:=(1-t)p+tq$ and $\tau_t:=\sigma_t^{\text{\tiny{T}}}$, Lemma \ref{lem:transpose} makes $\tau_t$ rejected, while separate affinity gives $U(\sigma_t)=U(\tau_t)=0$.

Fix $s\in\Delta$ and $t_0\in(0,1)$. Since $\check\Pi$ is open, $(1-\varepsilon)\tau_{t_0}+\varepsilon sp^{\text{\tiny{T}}}$ is rejected for all sufficiently small $\varepsilon>0$. Its first marginal is interior and its utility is $\varepsilon s^{\text{\tiny{T}}}(\phi+\psi)p$, so the interior representation implies $s^{\text{\tiny{T}}}(\phi+\psi)p<0$. On the other hand, openness of $\widehat\Pi$ implies that, for every $t\in(0,1)$ and all sufficiently small $\varepsilon>0$, $(1-\varepsilon)\sigma_t+\varepsilon sr_t^{\text{\tiny{T}}}$ is strict. Its first marginal is interior and its utility is $\varepsilon s^{\text{\tiny{T}}}(\phi+\psi)r_t$, so $s^{\text{\tiny{T}}}(\phi+\psi)r_t>0$ for every $t\in(0,1)$. This contradicts $r_t\to p$ as $t\downarrow0$.

Therefore a zero-utility distribution cannot be strict. It cannot be rejected either, because its transpose would then be strict by Lemma \ref{lem:transpose} and would also have zero utility. It must be indifferent, completing the  extension.
\end{proof}

\begin{proof}[Proof of Lemma \ref{lem:cwu-ceu}]
Suppose first that $\phi\notin\Sspace=\Tspace^\perp$. Choose $Z\in\Tspace$ with $\langle\phi,Z\rangle=1$, fix $r\in\Delta$, and write $A:=\phi+\psi$. For $p,q$ sufficiently close to $r$, let $\tau_{pq}:=pq^{\text{\tiny{T}}}-(p^{\text{\tiny{T}}}Aq)Z$. Because $A$ is skew-symmetric, $r^{\text{\tiny{T}}}Ar=0$; hence, after shrinking the neighborhood of $r$, $\tau_{pq}$ is nonnegative. Its marginals are $(p,q)$, and $U(\tau_{pq})=p^{\text{\tiny{T}}}\phi q-p^{\text{\tiny{T}}}Aq+
p^{\text{\tiny{T}}}\psi q=0$.

Thus all such $\tau_{pq}$ are indifferent. Strong independence (i) implies that $\alpha\tau_{pq}+(1-\alpha)\tau_{p'q'}$ is indifferent for nearby $p,p',q,q'$ and $\alpha\in(0,1)$. Evaluating its CWU utility and using
$U(\tau_{pq})=U(\tau_{p'q'})=0$ yields \(
 (p-p')^{\text{\tiny{T}}}\psi(q-q')=0
\). Nearby lottery differences span $H:=\{d\in\R^n:\one^{\text{\tiny{T}}}d=0\}$, whose rank-one products span $\Tspace$. Hence $\psi\perp\Tspace$, so $\psi\in\Sspace$.

It remains to consider $\phi\in\Sspace$. Put $A:=\phi+\psi$. Observe that $\langle\phi,\pi\rangle=\pi_1^{\text{\tiny{T}}}\phi\pi_2$, and hence $U(\pi)=\pi_1^{\text{\tiny{T}}}A\pi_2$. Fix $s\in\Delta$ and define $\ell(h):=h^{\text{\tiny{T}}}As$ on $H$. We first claim that $ h\in\ker\ell \implies  h^{\text{\tiny{T}}}A=0$.

Indeed, choose $\varepsilon>0$ small enough that $p:=s+\varepsilon h\in\Delta$. Then
$p^{\text{\tiny{T}}}As=0$, so every joint distribution with marginals $(p,s)$ is indifferent. For any $r\in\Delta$, mix one such distribution with the symmetric distribution $rr^{\text{\tiny{T}}}$. Strong independence (i) makes the mixture indifferent. Its marginals are $p_\alpha=q_\alpha+\alpha\varepsilon h$ and $q_\alpha=\alpha s+(1-\alpha)r$, so skew symmetry gives $0=p_\alpha^{\text{\tiny{T}}}Aq_\alpha
=\alpha\varepsilon h^{\text{\tiny{T}}}Aq_\alpha$. Letting $\alpha\downarrow0$ yields $h^{\text{\tiny{T}}}Ar=0$ for every $r\in\Delta$, proving the claim.

If $\ell=0$, the claim immediately gives $d^{\text{\tiny{T}}}Ae=0$ for all $d,e\in H$. Otherwise choose $k\in H$ with $\ell(k)=1$ and write $d=d_0+\ell(d)k$ and $e=e_0+\ell(e)k$, where $d_0,e_0\in\ker\ell$. Terms involving $d_0$ or $e_0$ vanish by the claim and skew symmetry, while $k^{\text{\tiny{T}}}Ak=0$. Thus again $d^{\text{\tiny{T}}}Ae=0$ on $H\times H$, so $A\in\Tspace^\perp=\Sspace$. Since also $\phi\in\Sspace$, $\psi=A-\phi\in\Sspace$, as required.
\end{proof}
\begin{proof}[Proof of Proposition \ref{prop:ceu}]
The ``if'' direction follows from linearity and skew symmetry. Conversely, reflexivity and the third clause of strong independence imply completeness: if both $\pi$ and $\pi^{\text{\tiny{T}}}$ were rejected, their symmetric midpoint would be rejected, contrary to reflexivity. Strong independence gives the three betweenness clauses on the entire joint-distribution simplex. Applying the separation argument from Proposition \ref{prop:cbu} with that simplex in place of a fixed-marginal fiber therefore gives an affine functional whose positive, zero, and negative sets are exactly $\widehat\Pi$, $\widetilde\Pi$, and $\check\Pi$. Since joint distributions have unit mass, represent it as $\langle\phi,\pi\rangle$ with threshold zero. Reflexivity makes every symmetric joint distribution have value zero. Evaluating at $\delta_{(x,x)}$ and at $\frac12\delta_{(x,y)}+\frac12\delta_{(y,x)}$ gives $\phi(x,x)=0$ and $\phi(x,y)+\phi(y,x)=0$ for every $x,y\in X$. Thus $\phi$ is skew-symmetric and gives the claimed CEU representation.
\end{proof}

\begin{proof}[Proof of Proposition \ref{prop:kis-sml-ci}]
Fix $d\ne0$ and set $F_d(\mathbf y):=\tilde u(d;\mathbf y)-\tilde u(0;\mathbf y)$. SML-CI gives $\mathbb E_P[F_d(\mathbf Y)]=0$ throughout $\Gamma_D(p^*,\ldots,p^*)$. Apply the multi-marginal version of Lemma~\ref{lem:transport-separable}. Specifically, the OT problem maximizing this expectation has value zero, and the product coupling $P^*:=(p^*)^{\otimes K}$ is a full-support optimum. Its Kantorovich dual minimizes $\sum_{k\in D}\sum_{x\in X}a_{dk}(x)p^*(x)$ subject to $\sum_{k\in D}a_{dk}(y_k)\geq F_d(\mathbf y)$ for every $\mathbf y\in X^D$. Finite-dimensional duality gives optimal potentials with objective value zero. The nonnegative dual slack therefore has zero expectation under $P^*$. Since $P^*(\mathbf y)>0$ at every vector, every slack is zero, yielding $F_d(\mathbf y)=\sum_{k\in D}a_{dk}(y_k)$. The representation follows by applying this argument to each $d\ne0$, and taking $h(\mathbf y):=\tilde u(0;\mathbf y)$ and $a_{0k}:=0$. The common expectation $\mathbb E_P[h(\mathbf Y)]$ cancels from all comparisons, giving global CI.
\end{proof}
\begin{proof}[Proof of Proposition \ref{prop: CDU CI}]
    Writing $m(p):=\int_0^1 h(u)d\mu_p(u)$ gives
\[
 U_\phi(\pi)=D_f(p)-D_f(q)+m(p)m(q)\varphi(c_\pi).
\]
For the uniform marginal $p^0$, $Q_{p^0}(u)=1-u$ and $m(p^0)=0$, so the preference is CI at $p^0$ against every $q$. For the atomless full-support marginal $p^1$ with $Q_{p^1}(u)=(1-u)^2$, one has $d\mu_{p^1}(u)=2(1-u)du$ and $m(p^1)=-1/3$. Its SML comparisons have value $\varphi(c_\pi)/9$, which takes opposite signs at a copula with nonzero premium and its transpose. Thus local insensitivity coexists with sensitivity elsewhere.
\end{proof}

\section{Alternative Proofs for SSB and WU}\label{appendix: SSB}\label{appendix:ssb}\label{appendix:rank}

\begin{proof}[Proof of Proposition \ref{lem:ssb-branch}]
By Proposition \ref{prop:pointwise-ci}, every local kernel  $\phi_p$ is separable for interior $p$. Write $ \phi_p=u_p\one^{\text{\tiny{T}}}-\one w_p^{\text{\tiny{T}}}$ and set $v_p:=(u_p\cdot p)\one-w_p$. For every $q$ and every $\pi\in\Gamma(p,q)$, $ \langle\phi_p,\pi\rangle=v_p\cdot q$. Thus $p\succeq^*q$ if and only if $v_p\cdot q\geq0$, with equality if and only if $p\sim^*q$, and reflexivity gives $v_p\cdot p=0$. This is a conditional linear utility representation \citep{fishburn1982nontransitive,nakamura1990bilinear}.
Continuity and correlation insensitivity make $\sim^*$ closed on $\Delta\times\Delta$, as seen by taking limits of the corresponding product couplings.

Construct the nonindifference graph on $\Delta$: connect $p$ and $q$ when $p\not\sim^*q$. Completeness makes $v_p\cdot q$ and $v_q\cdot p$ have opposite signs on an edge. Define
\[
 a_{pq}:=-\frac{v_p\cdot q}{v_q\cdot p}>0,
 \qquad a_{qp}=a_{pq}^{-1}.
\]
Following \citet{chew1994projective}, we show that projective independence implies the projective cycle identity. For three pairwise connected and affinely independent points $p,q,r$, restrict their indifference hyperplanes to the affine plane that they span. The planar consequence of projective independence established in Lemma \ref{lem:pi-pencil} below implies that these three lines are concurrent or parallel. In the concurrent case, write their common intersection in barycentric coordinates as $s=\alpha p+\beta q+\gamma r$, where the coefficients may extend beyond the simplex. Since $v_p\cdot p=0$ and $v_p\cdot s=0$, \(
 \frac{v_p\cdot q}{v_p\cdot r}=-\frac{\gamma}{\beta}
\), and the two analogous identities give
\[
 \frac{v_p\cdot q}{v_q\cdot p}
 \frac{v_q\cdot r}{v_r\cdot q}
 \frac{v_r\cdot p}{v_p\cdot r}=-1.
\]
In the parallel case, choose a common nonzero normal $n$ to the three lines and write $v_s \cdot x=k_s n \cdot (x-s)$ for $s \in \{p,q,r\}$. Then $a_{pq}=k_p/k_q$, and the cycle product is one.  If $p,q,r$ are affinely dependent, parameterize their common line by a scalar $t$. The restriction of each nonzero functional has the form $v_p\cdot x=k_p(t_x-t_p)$, so $a_{pq}=k_p/k_q$ and the product identity follows directly. Therefore $a_{pq}a_{qr}a_{rp}=1$ in every case. Two-cycles also have product one.

Every cycle of the nonindifference graph decomposes into two- and three-cycles. Given the finitely many vertices of a cycle, choose an auxiliary lottery outside the union of their proper indifference hyperplanes. It is connected to every vertex, so the original cycle is the product of the resulting triangular cycles, with the auxiliary edges cancelling in opposite orientations. If $v_p$ vanishes identically, then $p$ is indifferent to every lottery and is an isolated node; assign it an arbitrary positive scale. Consequently, every directed cycle $c$ in every nontrivial connected component satisfies $ \prod_{(p,q)\in c}a_{pq}=1$.

Choose one root in each component. The cycle identity gives a positive potential $\lambda$ satisfying $ a_{pq}=\lambda(q)/\lambda(p)$ on every edge. Define $\psi(p,q):=\lambda(p)v_p\cdot q$. On an edge, $ \lambda(p)v_p\cdot q=-\lambda(q)v_q\cdot p$; for an indifferent pair, both sides are zero. Hence $\psi(p,q)=-\psi(q,p)$ for every $p,q\in\Delta$. For a fixed $p\in\Delta$, $\psi(p,\cdot)$ is linear. Skew symmetry writes $\psi(\cdot,q)=-\lambda(q)v_q\cdot(\cdot)$, so it is linear in the first argument as well. Therefore $\psi$ is bilinear on $\Delta\times\Delta$ and has the form $\psi(p,q)=p^{\text{\tiny{T}}}\psi q$ for a skew-symmetric matrix, also denoted by $\psi$. For a fixed $p$, the two linear functions $q\mapsto p^{\text{\tiny{T}}}\psi q$ and $q\mapsto\lambda(p)v_p\cdot q$ agree on the relatively open set $\Delta$ and hence as vectors; their equality extends to every $q\in\Delta X$. Positive rescaling and correlation insensitivity show that this matrix represents the original preference whenever the first marginal is interior. Lemma \ref{lem:cwu-boundary}, applied with correlation kernel zero, extends the representation to every joint distribution.
\end{proof}

\begin{lem}[CLU and Projective Independence]\label{lem:pi-pencil}
Suppose the induced comparisons $\succeq^*$ on $\Delta$ is continuous and complete. Assume that it admit a conditional linear representation, $p\succeq^*q\iff v_p\cdot q\geq0$ and $p\sim^*q\iff v_p\cdot q=0$, and satisfies projective independence. For any pairwise non-indifferent and affinely independent $p,q,r\in\Delta$, the restrictions of their indifference hyperplanes to $A:=\operatorname{aff}\{p,q,r\}$ are concurrent or parallel.
\end{lem}

\begin{proof}
Fix such $p,q,r$, put $D:=A\cap\Delta$, and, for each $s\in D$, let $L_s:=\{t\in A:v_s\cdot t=0\}$. Reflexivity gives $s\in L_s$. If some $s\in D$ is indifferent to every point of $D$, symmetry puts $s$ on every nontrivial $L_t$, so the lines are concurrent. Otherwise every $L_s$ is a line. If distinct $s,t\in D$ are indifferent, reflexivity and symmetry give $L_s=L_t$. Continuity makes $\sim^*$ closed. The line direction varies continuously with $s$. To see this, let $s_m\to s$ and unit directions $d_m$ of $L_{s_m}$ converge to $d$. Choose $\varepsilon>0$ small enough such that $s_m+\varepsilon d_m\in D$ for all sufficiently large $m$. Closedness then gives $s\sim^*s+\varepsilon d$, so $d$ is a direction of $L_s$. 

Choose local coordinates in which the lines are nonvertical. On a transverse segment $P_y=(0,y)$, write the direction of $L_{P_y}$ as $(1,f(y))$, where $f$ is continuous. After shrinking the neighborhood, every indifference line meets this segment and equals the line indexed by its intersection point. For endpoints $y_0,y_1$, scalars $c_0,c_1$, and sufficiently small $\varepsilon>0$, the points $Q_i:=P_{y_i}+\varepsilon c_i(1,f(y_i))$ lie in $D$ and satisfy $P_{y_i}\sim^* Q_i$. With $y_t:=ty_0+(1-t)y_1$, their $t$-mixtures are indifferent exactly when
\[
 t c_0\{f(y_0)-f(y_t)\}+(1-t)c_1\{f(y_1)-f(y_t)\}=0.
\]
If $f(y_0)\neq f(y_1)$, fix $\alpha\in(0,1)$ and choose a nonzero $(c_0,c_1)$ satisfying this equality at $t=\alpha$. Projective independence makes it hold for every $t\in(0,1)$. Neither coefficient can be zero, since otherwise $f$ would equal one endpoint value throughout the open segment, contradicting continuity at the other endpoint. Therefore
\[
 f(y_t)=\frac{tc_0f(y_0)+(1-t)c_1f(y_1)}{tc_0+(1-t)c_1}.
\]
The denominator cannot vanish, since the two displayed equalities would then imply $tc_0\{f(y_0)-f(y_1)\}=0$. Hence $f$ is locally either constant or fractional linear.

Move the transverse line to $x=\delta$. If $f_\delta$ is its slope function, points on the same indifference line satisfy $f_\delta(y+\delta f(y))=f(y)$. In the nonconstant case, shrink the interval and choose $\delta\neq0$ so that $g(y):=y+\delta f(y)$ is invertible. Both $f$ and $f_\delta$ are nonconstant fractional-linear functions, so $g=f_\delta^{-1}\circ f$ is fractional linear. Writing $f(y)=(ay+b)/(cy+d)$ with $ad-bc\neq0$ gives
\[
 g(y)=\frac{cy^2+(d+\delta a)y+\delta b}{cy+d}.
\]
If $c\neq0$, the numerator and denominator have no common root, because the numerator at $-d/c$ equals $\delta(bc-ad)/c\neq0$; hence $g$ cannot be fractional linear. Thus $c=0$ and $f(y)=my+b_0$ is affine. The local lines pass through $(-1/m,-b_0/m)$ when $m\neq0$ and are parallel when $m=0$. Local pencils agree on open overlaps, since two distinct pencils have at most one common line. Connectedness of $D$ therefore gives one pencil on all of $D$.
In particular, $L_p,L_q,L_r$ are concurrent or parallel.
\end{proof}

The following rank-two lemma underpins the reduction of SSB to WU.
\begin{lem}[Rank Two]\label{lem:rank-two}
Let $\psi\in\R^{n\times n}$ be skew-symmetric and $\Psi(p,q):=p^{\text{\tiny{T}}}\psi q$. If for all nonzero $p,q,r\in\R^n_+$, $\Psi(p,q)=0=\Psi(q,r)\implies\Psi(r,p)=0$, then $\operatorname{rank}(\psi)\leq2$.
\end{lem}

\begin{proof}
If $\psi=0$, the conclusion is immediate. Otherwise choose $q\gg0$ such that $\psi q\neq0$, which is possible because the kernel of a nonzero matrix is a proper subspace. Let $H_q:=\{x:x^{\text{\tiny{T}}}\psi q=0\}$ and $C_q:=H_q\cap\R^n_+$. Skew symmetry gives $q\in C_q$, so set $z=q\gg0$.

We first show that $\Psi$ vanishes on $H_q\times H_q$. For $p,r\in C_q$, one has $\Psi(p,q)=0=\Psi(q,r)$, so the hypothesis gives $\Psi(r,p)=0$. Let $a,b\in H_q$. For sufficiently small $\varepsilon>0$, all four vectors $z\pm\varepsilon a$ and $z\pm\varepsilon b$ belong to $C_q$. Applying the preceding conclusion to the four pairs and taking the alternating sum gives $4\varepsilon^2\Psi(a,b)=0$. Thus $\Psi(a,b)=0$ for all $a,b\in H_q$.

For $a\in H_q$ and every $b\in H_q$, the identity $b^{\text{\tiny{T}}}\psi a=0$ implies $\psi a\in H_q^\perp=\operatorname{span}\{\psi q\}$. Hence $\psi(H_q)$ has dimension at most one. Choose $u\notin H_q$. Since $\R^n=H_q\oplus\operatorname{span}\{u\}$, $ \psi(\R^n)\subseteq\operatorname{span}\{\psi q,\psi u\}$, so $\operatorname{rank}(\psi)\leq2$. The rank of a skew-symmetric matrix is even.
\end{proof}

\begin{proof}[Proof of Proposition \ref{prop:ssb-wu}]
By positive homogeneity, CI-transitivity on lotteries implies the indifference implication in Lemma \ref{lem:rank-two} on all nonzero vectors in $\R^n_+$, so $\operatorname{rank}(\psi)\leq2$. The zero-rank case is complete indifference and is WU with $a=0$ and $w=\one$. Suppose the rank is two and write $\psi=aw^{\text{\tiny{T}}}-wa^{\text{\tiny{T}}}$. Completeness and CI-transitivity order the degenerate lotteries. Since $\psi\neq0$, choose outcomes $x_-$ and $x_+$ that are respectively worst and best, with $\delta_{x_+}\succ^*\delta_{x_-}$. Then $(a(x_-),w(x_-))$ and $(a(x_+),w(x_+))$ are non-collinear. These two vectors lie in a common open half-plane, so a non-singular two-by-two change of basis with positive determinant can be chosen with $w(x_-),w(x_+)>0$; it changes $\psi$ only by a positive scalar and preserves $a(x_+)w(x_-)-w(x_+)a(x_-)>0$. For every intermediate outcome $x$,
\[
 w(x)=\frac{w(x_-)[a(x_+)w(x)-w(x_+)a(x)]+w(x_+)[a(x)w(x_-)-w(x)a(x_-)]}{a(x_+)w(x_-)-w(x_+)a(x_-)}>0,
\]
because $a(x_+)w(x)-w(x_+)a(x)\geq0$ and $a(x)w(x_-)-w(x)a(x_-)\geq0$ express that $\delta_{x_+}\succeq^*\delta_x\succeq^*\delta_{x_-}$. At least one inequality is strict: equality in both would make $\delta_x$ indifferent to both $\delta_{x_+}$ and $\delta_{x_-}$, contradicting $\delta_{x_+}\succ^*\delta_{x_-}$. Thus $p^{\text{\tiny{T}}}\psi q$ has the same sign as $(a^{\text{\tiny{T}}}p)/(w^{\text{\tiny{T}}}p)-(a^{\text{\tiny{T}}}q)/(w^{\text{\tiny{T}}}q)$, which is WU. The converse and the further reduction from WU to EU under ordinary independence are standard \citep{chew1983generalization}.
\end{proof}

\section{Proofs for the Finite-Support Extension}
\label{appendix:finite-support-proofs}

\begin{proof}[Proof of Theorem \ref{thm:support-lattice}]
Fix a finite $S$.  The restriction $\Pi_S$ is a correlation preference on
the ordinary finite simplex $\mathcal J(S)$ and inherits continuity,
completeness, and betweenness. Theorems \ref{thm:dichotomy} therefore gives exactly one of its two branches on $\Delta^\circ(S)$.  In the insensitive branch, the pointwise CI proposition
gives $S$-CI.  In the sensitive branch, the finite correlation-sensitive CWU
theorem extends the nondegenerate representation to all of
$\mathcal J(S)$. 

For monotonicity, let $S\in\mathfrak S$ and $S\subseteq T$.  Choose $p\in\Delta^\circ(S)$ and a strict
$\rho\in\Gamma_f(p,p)$.  Such a strict point exists because a nonindifferent point is either strict or has a strict transpose.  Choose $r\in\Delta^\circ(T)$ and define, for $\varepsilon\in(0,1)$, $p_\varepsilon:=(1-\varepsilon)p+\varepsilon r$ and $ \rho_\varepsilon:=(1-\varepsilon)\rho+\varepsilon(r\otimes r)$. Then $\rho_\varepsilon\in\Gamma_f(p_\varepsilon,p_\varepsilon)$ and $p_\varepsilon\in\Delta^\circ(T)$.  Since $\widehat\Pi$ is open and $\rho_\varepsilon\to\rho$, $\rho_\varepsilon$ is strict for all sufficiently small $\varepsilon>0$.  Hence $T\in\mathfrak S$.  The second implication is its contrapositive.
\end{proof}

\begin{lem}
\label{lem:finite-cwu-uniqueness}
Let $F\subset X$ be finite. Suppose $U_i(\pi)=\ip{\phi_i}{\pi}+\pi_1^{\T}\psi_i\pi_2$, $i=1,2$, are two CWU representations of the same preference on $\mathcal J(F)$, and suppose the preference is SML-sensitive at an interior marginal. Then there is a scalar $c>0$ such that $U_2=cU_1$ on $\mathcal J(F)$.
\end{lem}

\begin{proof}
The finite dichotomy implies that every $p\in\Delta^\circ(F)$ is SML-sensitive.  On the fixed-first-marginal fiber $K_{p,F}:=\{\pi\in\mathcal J(F):\pi_1=p\}$, both $U_1$ and $U_2$ are nonconstant affine functions with identical positive, zero, and negative sets. Their common zero set contains $pp^{\text{\tiny{T}}}$, which lies in the relative interior of $K_{p,F}$, and is the intersection of that fiber with an affine hyperplane. Equality of the zero sets therefore makes their affine normals proportional; equality of the positive sides makes the proportionality factor positive. Consequently, there is a unique $c(p)>0$ such that $ U_2(\pi)=c(p)U_1(\pi)
 $ for $\pi\in K_{p,F}$. SML sensitivity provides a matrix $Z$ on $F^2$ with zero row and column sums and
$\ip{\phi_1}{Z}\neq0$.  For every interior $p$, sufficiently small positive and negative perturbations of $p\otimes p$ in direction $Z$ are feasible, so that $\ip{\phi_2}{Z}=c(p)\ip{\phi_1}{Z}$. Hence $c(p)$ is a common constant $c$ on $\Delta^\circ(F)$.  We have $U_2=cU_1$ whenever the first marginal is interior.  Both sides are finite-dimensional polynomial functions of the entries of $\pi$, and interior-first-marginal distributions are dense in $\mathcal J(F)$.  The identity therefore extends to the whole face.
\end{proof}

\begin{proof}[Proof of Theorem \ref{thm:direct-limit-cwu}]
\emph{Necessity.} Assume the three axioms and let $p^0$ be SML-sensitive.  Put $S_0:=\supp(p^0)$.  Pick a $\bar\pi\in\Gamma_f(p^0,p^0)$ in the strict preference set. Let $F\supseteq S_0$ be finite. By the monotonicity argument in Theorem~\ref{thm:support-lattice}, the restriction to $\mathcal J(F)$ has an SML-sensitive interior marginal.  The finite correlation-sensitive CWU theorem therefore supplies a nondegenerate CWU functional $U_F$ representing the preference on all of $\mathcal J(F)$.  Rescale it so that $U_F(\bar\pi)=1$.

These normalized functionals agree on overlaps. Indeed, if $F,G\supseteq S_0$ and $H:=F\cap G$, then $H\supseteq S_0$ and hence $H\in\mathfrak S$.  The restrictions of $U_F$ and $U_G$ to $\mathcal J(H)$ are exact CWU representations of the same sensitive-face preference.  Lemma~\ref{lem:finite-cwu-uniqueness} makes them positively proportional, and the scaling earlier makes the factor one. Consequently, $ U_F|_{\mathcal J(H)}=U_G|_{\mathcal J(H)}$.

It remains to choose compatible decompositions into correlation and marginal
parts.  Fix $x_0\in S_0$.  Write $ U_F(\pi)=\ip{\phi_F}{\pi}
 +\pi_1^\T\psi_F\pi_2$ with skew matrices $\phi_F,\psi_F$. Let
$a_F(x):=\phi_F(x,x_0)$ and apply the separable gauge transformation
\begin{align}
 \widetilde\phi_F(x,y)
 &:=\phi_F(x,y)-a_F(x)+a_F(y),\label{eq:canonical-gauge-phi}\\
 \widetilde\psi_F(x,y)
 &:=\psi_F(x,y)+a_F(x)-a_F(y).
 \label{eq:canonical-gauge-psi}
\end{align}
This leaves $U_F$ unchanged and imposes $\widetilde\phi_F(x,x_0)=0$ for every $x\in F$.

The canonical decompositions also agree on overlaps. Suppose two skew pairs $(\phi,\psi)$ and $(\phi',\psi')$ produce the same total functional on $\mathcal J(H)$. Perturbing a full-support product distribution by sufficiently small zero-marginal matrix $Z$ gives $\ip{\phi-\phi'}{Z}=0$. Finite transport separability therefore implies $\phi-\phi'=b\one^\T-\one b^\T$ for some vector $b$. If both correlation kernels vanish on $H\times\{x_0\}$, then $b$ is constant and $\phi=\phi'$.  Evaluating at product measures then gives $\psi=\psi'$. This proves compatibility of \eqref{eq:canonical-gauge-phi}--\eqref{eq:canonical-gauge-psi}.

For $x,y\in X$, choose any finite $F$ containing $S_0\cup\{x,y\}$ and define $\phi(x,y):=\widetilde\phi_F(x,y)$, $ \psi(x,y):=\widetilde\psi_F(x,y)$. Compatibility makes these definitions independent of $F$.  They are skew. Define, on finitely supported signed measures,
\begin{align*}
 L(Z):=\sum_{x,y}\phi(x,y)Z(x,y),\quad
 B(r,s):=\sum_{x,y}\psi(x,y)r(x)s(y).
\end{align*}
The sums are finite, $L$ is skew linear, and $B$ is skew bilinear.  Given any $\pi\in\Jf(X)$, choose a finite $F$ containing $S_0$ and both marginal supports.  Construction gives $ L(\pi)+B(\pi_1,\pi_2)=U_F(\pi)$. Thus the CWU represents the preference on all of $\Jf(X)$, including faces disjoint from $S_0$.  It is nondegenerate because
its restriction to $\mathcal J(S_0)$ is nondegenerate. Its nonnegative set is $\Pi$, which is weakly closed by continuity.

\noindent\emph{Sufficiency.} Conversely, let $(L,B)$ be a weakly closed, nondegenerate functional CWU.
Weak closedness gives continuity.  Skew symmetry gives $U(\pi^\T)=-U(\pi)$ which implies completeness and identifies positive, zero, and negative values with strict preference, indifference, and rejection.  If the first marginal is fixed at $p$, then $ U(\alpha\pi+(1-\alpha)\pi')
 =\alpha U(\pi)+(1-\alpha)U(\pi')$ because $L$ is linear and $B(p,\cdot)$ is linear.  The three clauses of betweenness follow.

Finally, nondegeneracy supplies a finitely supported signed measure $Z$ with $Z_1=Z_2=0$ and $L(Z)\neq0$.  Let $S$ contain all row and column coordinates in $\supp(Z)$, and choose $p\in\Delta^\circ(S)$.  For sufficiently small $\varepsilon>0$, $ \pi^\pm:=p\otimes p\pm\varepsilon Z$ are nonnegative and belong to $\Gamma_f(p,p)$.  Since $p\otimes p$ is symmetric and $B(p,p)=0$, $ U(\pi^\pm)=\pm\varepsilon L(Z)$. The two couplings have opposite signs, so $p$ is SML-sensitive.
\end{proof}

\end{document}